\documentclass[12pt]{iopart}

\usepackage{iopams}

\usepackage[american]{babel}
\usepackage{amsfonts}
\usepackage{amssymb}
\usepackage{epsfig}
\usepackage{latexsym}
\usepackage{paralist}
\usepackage{fancyhdr}
\usepackage{graphicx}
\usepackage{graphicx,txfonts}
\usepackage{multirow}
\usepackage{placeins}

\usepackage[symbol]{footmisc}
\usepackage{etex}
\usepackage{braket}
\usepackage{float}
\usepackage{slashed}
\usepackage{xcolor}
\usepackage{soul}
\usepackage{dcolumn} 
\usepackage{bm}
\usepackage{nicefrac}
\usepackage{tipa}
\usepackage{dsfont}
\usepackage{caption}

\def\correspondingauthor{\footnote{Corresponding Author}}

\begin{document}

\title[Covariant spin-parity decomposition of the Torsion and Path Integrals]{Covariant spin-parity decomposition of the Torsion and Path Integrals}

\author{Riccardo Martini}
\address{INFN - Sezione di Pisa, Largo Bruno Pontecorvo 3, 56127 Pisa, Italy}
\ead{riccardo.martini@pi.infn.it}

\author{Gregorio Paci}
\address{
	Universit\`a di Pisa and INFN - Sezione di Pisa, Largo Bruno Pontecorvo 3, 56127 Pisa, Italy}
	\ead{gregorio.paci@phd.unipi.it}

\author{Dario Sauro\correspondingauthor}
\address{
	Universit\`a di Pisa and INFN - Sezione di Pisa, Largo Bruno Pontecorvo 3, 56127 Pisa, Italy}
\ead{dario.sauro@phd.unipi.it}

\vspace{0.5cm}

\begin{indented}
\item[]Received 11 May 2023; revised 21 July 2023
\item[]Accepted for publication 15 August 2023
\item[]Published 30 August 2023

\end{indented}

\begin{abstract}
	We propose a functional measure over the torsion tensor. We discuss two completely equivalent choices for the Wheeler-DeWitt supermetric for this field, the first one is based on its algebraic decomposition and the other is inspired by teleparallel theories of gravity. The measure is formally defined by requiring the normalization of the Gau{\ss}ian integral. To achieve such a result we split the torsion tensor into its spin-parity eigenstates by constructing a new, York-like, decomposition. Of course, such a decomposition has a wider range of applicability to any kind of tensor sharing the symmetries of the torsion. As a result of this procedure a functional Jacobian naturally arises, whose formal expression is given exactly in the phenomenologically interesting limit of maximally symmetric spaces. We also discuss the explicit computation of this Jacobian in the case of a $4$-dimensional sphere $S^4$ with particular emphasis on its logarithmic divergences.
\end{abstract}

\vspace{3cm}

\CQG \,\, \textbf{40} \,\, 195005

\maketitle

\section{Introduction}

It is commonly accepted that gravity can be classically described as a dynamical theory of spacetime geometry. This point of view was first established in Einstein's groundbreaking work \cite{Einstein:1916vd} that led to the formulation of general relativity (GR), with the metric tensor playing the role of the fundamental field variable.

Although it is elegantly described by differential geometry, GR suffers from some flaws at both the quantum and classical levels. For example, it is well known to be not perturbatively renormalizable in four dimensions \cite{tHooft:1974toh,Christensen:1979iy,Goroff:1985th} as well as to display singularities with great generality \cite{Raychaudhuri:1953yv,Penrose:1964wq,Hawking:1970zqf,Hawking:1973uf}. This situation points toward the necessity to consider a more generic theoretical framework. As already considered by Einstein himself shortly after developing GR, one of the most natural ideas to take into account is to allow for more general geometries. For example, one can study the geometry of a smooth manifold that is not uniquely described by the metric tensor, i.e., that need not be Riemannian. These kinds of geometries are commonly dubbed metric-affine geometries, or metric-affine theories of gravity (MAG) in the physics literature \cite{Kibble:1961ba,Hehl:1976kj,Hehl:1994ue,Gronwald:1995em,Sezgin:1979zf,Sezgin:1981xs,Capozziello:2011et,BeltranJimenez:2019esp,BeltranJimenez:2019acz,BeltranJimenez:2020sqf,BeltranJimenez:2020sih,Sauro:2022chz,Sauro:2022hoh,Percacci:2023rbo}. The post-Riemannian fields which, together with the metric, describe these geometries are the torsion and the non-metricity. The first is responsible for the non-closure of infinitesimal parallelograms, while the latter shears and dilates the vectors that are parallel transported around closed loops.

In GR both torsion and non-metricity are set to zero kinematically. However, these kinematical constraints can be derived dynamically from the Hilbert action in a more generic setting. Indeed,
we can impose zero torsion with the field equations for the spin connection, which can be derived by means of exterior calculus \cite{Dadhich:2012htv}, while the condition of zero non-metricity is equivalent to the equations of motion for a holonomic symmetric affine-connection {\cite{Palatini:1919}. There are two main ways to understand how torsion arises through gauge-theoretic methods. On the one hand one, has the Poincar\'e gauge theories, where the co-frame is the affine potential of translations and the torsion is its curvature $2$-form \cite{Hehl:1976kj,Hehl:1994ue}. On the other hand, one can start from a general MAG where the full general linear group on the tangent bundle is gauged. As it is well known, torsion is tightly connected with the Lorentz subgroup of $GL(4,\mathbb{R})$ (see \cite{Blagojevic:2013xpa} for a comprehensive review on gravity as a gauge theory),} while non-metricity is tied to its complement \cite{Hehl:1994ue,Gronwald:1995em}. We know that the Lorentz group is a symmetry of nature, so we require this symmetry to be realized as a gauge symmetry in the high-energy regime, whereas we do not ask for invariance under the full local $GL(4,\mathbb{R})$.  In addition, notice that, for generic Lagrangians, a non-vanishing non-metricity would also imply the propagation of a potentially dangerous spin-$3$ particle \cite{Percacci:2020ddy,Baldazzi:2021kaf,2647462,Weinberg:1965nx}. On the other hand, the spin-parity content of the torsion spans all the possible parity states with spin $J=0,1,2$ \cite{Sezgin:1981xs,Baldazzi:2021kaf}. For these reasons, in this work we retain the torsion while discarding the non-metricity. However, it is important to notice that we can choose the metric gauge \cite{Percacci:2020bzf} by trivializing the co-frame. This way, we have a gravitational theory described only by a metric field and by an independent connection. In this paper we carry out computations in the metric gauge, where the invariance under the general linear group is not manifest, {in fact, it is completely broken. As consequence of this gauge choice, there is no difference between tangent and fiber bundle indices. Lorentz invariance can be restored by choosing the tetrad gauge instead of the metric one \cite{Percacci:2020bzf}. Our choice of gauge is only due to computational convenience.}

Even though torsionful theories have been studied for a long time \cite{Kibble:1961ba,Hehl:1976kj,Sezgin:1979zf,Sezgin:1981xs,Shapiro:2001rz}, only recently a systematic analysis of their quantum properties has been undertaken for completely general models, albeit only in the flat-space limit \cite{Baldazzi:2021kaf}. This is surely due to the huge complexity of these theories, which involve a very large number of \emph{a priori} independent coupling constants \cite{Baldazzi:2021kaf}. Indeed, previous analyses \cite{Sezgin:1979zf,Sezgin:1981xs,Percacci:2020ddy} have started from the simplifying assumption that the full action is quadratic in either the full curvature or the torsion, thus reducing the number of independent coupling constants to six. {However, even though there are some arguments for imposing such a requirement \cite{Fabbri:2014dxa},} it has not been proven that this assumption is stable against quantum fluctuations. Moreover, recently some of us have shown that this guiding principle is incompatible with the conformal coupling of the trace-free hook antisymmetric part of the torsion in a Riemannian background \cite{Paci:2023twc}. Therefore, we feel the urge to have new covariant computational tools for tackling the study of the quantum properties of torsionful geometries on non-trivial Riemannian backgrounds.

Both in constrained systems and un-constrained ones it is highly convenient to decompose tensor fields into their spin-parity components. At least at the linearized level, this procedure highlights the physical content of the theory and simplifies considerably the task of writing down the propagators for a given kinetic term \cite{Percacci:2017fkn}. This kind of decomposition is well known in flat space, whereas we are only aware of some extensions to curved spaces, most notably the Hodge decomposition of $p$-forms on Riemannian manifolds \cite{Hodge:1989,Warner:1983fdm} and the York decomposition of symmetric rank-$2$ tensors \cite{Deser:1967zzb,York:1973ia,York:1974psa}. The usefulness of a covariant spin-parity decomposition lies in the geometric definition of the functional integral \cite{Mazur:1990ak,Mottola:1995sj}, which allows us to integrate over transverse fields only, and to circumvent the complicated covariant techniques needed when non-minimal operators are present \cite{Barvinsky:1985an}. In this paper we address this problem for the torsion tensor, i.e., we provide a covariant spin-parity decomposition of the torsion tensor which is in principle valid in every dimension. Our decomposition is tailored for $1$-loop calculations of torsion fluctuations in a Riemannian background. Non-trivial torsion backgrounds would necessitate a more complicated analysis which is beyond the scope of this paper.

From a dynamical point of view, the spin-parity decomposition can serve as a guide to identify the action of possible gauge symmetries that would render (some of) the longitudinal modes unphysical. Although this is not a necessary condition, requiring the dynamics to obey such symmetries has a twofold advantage: first, the perspective according to which the longitudinal modes decouple as the effect of a transformation (rather than just a decomposition) lifts the zero-modes of the differential operators that we study in the following to be generators of actual symmetries of background solutions. It is then natural to wonder what would be the gauge counterpart of such transformations and what constraints they would impose on the dynamics of torsion fluctuations. Second, from a computational point of view, one could exploit the consequent gauge symmetry to get rid of the non-minimal terms in the kinetic operator, therefore simplifying a lot the implementation of covariant computations based on heat kernel methods.

The structure of the paper is as follows. In Sect.\ \ref{sect:IntroMag} we provide a very brief introduction to metric-affine theories, emphasizing the reasons that lead us to discard the non-metricity.
	
In Sect.\ \ref{sect:Transv} we decompose the torsion tensor $T^\rho{}_{\mu\nu}$ into its completely antisymmetric $H^\rho{}_{\mu\nu}$ and hook antisymmetric $t^\rho{}_{\mu\nu}$ parts. The latter is analyzed first, and divided into its transverse and longitudinal modes. These are parametrized by a symmetric $S^{\mu\nu}$ and an antisymmetric $A_{\mu\nu}$ $2$-tensors through first-order linear differential operators, which have the same tensor structure as Curtright's gauge transformations \cite{Curtright:1980yk}. These tensors can be found by inverting the coupled system of second-order elliptic operators found by computation of the two algebraically independent divergences of $t^\rho{}_{\mu\nu}$. In particular, from $(div_1 t)_{\mu\nu} \equiv \nabla_\rho t^\rho{}_{\mu\nu}$ we invert the second-order operator to find $A_{\mu\nu}$, while from $(div_2 t)^{\mu\nu} \equiv \nabla_\rho (t^{\mu\rho\nu}+t^{\nu\rho\mu})$ we solve for $S^{\mu\nu}$. Zero-modes of the first-order differential operators, which we shall call the Curtright forms, and of the second-order differential operators are shown to be in one-to-one correspondence. Thus, they never contribute to the decomposition. The invertibility of these operators is guaranteed on closed manifolds since we find $A_{\mu\nu}$ and $S^{\mu\nu}$ to be orthogonal to their sources \cite{York:1974psa}. Eventually, we exploit the Hodge decomposition to divide the fully antisymmetric part $H^\rho{}_{\mu\nu}$ into transverse and longitudinal modes in a generic dimension $d$.

In section \ref{sect:TransvT} we further decompose $t^\rho{}_{\mu\nu}$ into its trace-free and pure-trace parts, which we are going to denote $\kappa^\rho{}_{\mu\nu}$ and $\tau_\mu$, respectively. To maintain the tracelessness while singling out the transverse part of $\kappa^\rho{}_{\mu\nu}$, we write the trace-free versions of the Curtright forms. We check that the same mathematical properties required in the transverse case hold here as well, so that the two relevant second-order differential operators stemming from the trace-free Curtright forms are invertible too. Then, we split $A_{\mu\nu}$ and $S^{\mu\nu}$ into their transverse and longitudinal modes. The latter are described by two transverse vectors, which appear in the decomposition with the same tensor structure first written by Curtright. Only one of them is physical, as checked in the flat-space limit. This redundancy may be described through some kind of gauge transformations if we choose not to split $A_{\mu\nu}$ and $S^{\mu\nu}$ into their spin-parity eigenstates. We analyze shortly the zero-modes of the tracefree Curtright forms on maximally symmetric and flat spaces. In the first ones, the system of second-order differential operators decouples, while in the latter we find the two longitudinal zero-modes $A_{\mu\nu}$ and $S^{\mu\nu}$ to be harmonic functions. We close this section with a brief comparison of the two decompositions, which turns out to yield results analogous to those found by York \cite{York:1974psa}.

In section \ref{sect:WD&FM} we define a convenient parametrization for the Wheeler-DeWitt supermetric for the torsion. In contrast to the metric case, we now have two non-irrelevant constants, and we analyze for which values the supermetric is positive definite by computing the eigenvalues of the fully antisymmetric and pure-trace parts. A different parametrization of the supermetric is implicit in the well-known basis for scalars quadratic in torsion \cite{Kibble:1961ba,Sezgin:1979zf,Sezgin:1981xs,BeltranJimenez:2019acz,BeltranJimenez:2019esp,BeltranJimenez:2020sih,BeltranJimenez:2020sqf,Percacci:2020bzf,Percacci:2020ddy,Baldazzi:2021kaf}. We write down the explicit change of basis in $d=4$. Our choice has the virtue that the eigenvalue of the purely tensorial, top-spin, part $\kappa^\rho{}_{\mu\nu}$ is one (up to an irrelevant normalization constant), in close analogy with the supermetric for metric perturbations \cite{Mottola:1995sj}. Given the supermetric, we formally define the functional measure through the normalization of the Gau{\ss}ian integral. Then we plug the covariant transverse traceless decomposition of the torsion \eref{eq:TTdecomp-full}, which is one of our major results, into both the supermetric and the functional measure. The change of variables in field space generates a nontrivial Jacobian in the functional measure, whose expression is given exactly in the limit of the maximally symmetrical spaces \eref{eq:Jacobian-max-sym}. The logarithmically divergent terms of this Jacobian are computed on an $S^4$, mostly exploiting known results \cite{Codello:2008vh}.

\section{MAGs in a nutshell}\label{sect:IntroMag}

Contrary to GR, where the components of the connection are assumed to coincide with the Christoffel symbols, a generic connection $\Gamma^\rho{}_{\mu\nu}$ on the tangent bundle of a manifold $\{{\cal M}, g\}$ can be endowed with both non-metricity and torsion. In formulae, one can always parametrize the connection as
\begin{eqnarray}
\label{MAG-aff-conn-split}
\Gamma^\rho{}_{\nu\mu} = \mathring{\Gamma}^\rho{}_{\nu\mu} + \Phi^\rho{}_{\nu\mu} = \mathring{\Gamma}^\rho{}_{\nu\mu} + K^\rho{}_{\nu\mu} + N^\rho{}_{\nu\mu} \, ,
\end{eqnarray}
where $\mathring{\Gamma}^\rho{}_{\nu\mu}$ are the Christoffel symbols given by the metric $g_{\mu\nu}$, $\Phi^\rho{}_{\mu\nu}$ and $K^\rho{}_{\nu\mu}$ are known as the distortion and the contortion respectively, and we refer to $N^\rho{}_{\nu\mu}$ as the disformation. Note that, in our convention, the directional index of the connection is the rightmost one, i.e., for a vector field $u^\rho$ we have
\begin{eqnarray}
\nabla_{\mu}u^\rho = \partial_\mu u^\rho + \Gamma^\rho{}_{\nu\mu}u^\nu = \mathring{\nabla}_\mu u^\rho + \Phi^\rho{}_{\nu\mu}u^\nu\,.
\end{eqnarray}
Using the metric and its inverse to rise and lower spacetime indices, we have that the algebraic properties of the contortion and the disformation are
\begin{eqnarray}
K_{\rho\nu\mu} = - K_{\nu\rho\mu}\,,\qquad N_{\rho\nu\mu} = N_{\rho\mu\nu}\,,
\end{eqnarray}
and they can be traced back to the presence of torsion and non-metricity in the full connection. In fact, we can construct the following tensors out of the full connection
\begin{eqnarray}
	R^\lambda{}_{\rho\mu\nu} \equiv & \, \partial_\mu \Gamma^\lambda{}_{\rho\nu} - \partial_\nu \Gamma^\lambda{}_{\rho\mu} + \Gamma^\lambda{}_{\kappa\mu} \Gamma^\kappa{}_{\rho\nu} - \Gamma^\lambda{}_{\kappa\nu} \Gamma^\kappa{}_{\rho\mu} \,,\label{eq:riemannDef}\\
	T^\lambda{}_{\mu\nu} \equiv & \, \Gamma^\lambda{}_{\nu\mu} - \Gamma^\lambda{}_{\mu\nu} \,,\\
	Q_{\mu\nu\lambda} \equiv & \, - \nabla_\lambda g_{\mu\nu} \,,
\end{eqnarray}
and we have that
\begin{eqnarray}\label{def-contortion}
K^\rho{}_{\nu\mu} = \frac{1}{2} \left( T_\nu{}^\rho{}_\mu + T_\mu{}^\rho{}_\nu - T^\rho{}_{\nu\mu} \right) \, , \qquad  N^\rho{}_{\nu\mu} = \frac{1}{2} \left( Q_\nu{}^\rho{}_\mu + Q_\mu{}^\rho{}_\nu - Q_{\mu\nu}{}^\rho \right) \, .
\end{eqnarray}
The tensor $T^\lambda{}_{\mu\nu}$ is the torsion tensor while $Q_{\mu\nu\lambda}$ is the non-metricity. Note that, by construction, the post-Riemannian fields have the following algebraic properties  $T^\rho{}_{\mu\nu}=-T^\rho{}_{\nu\mu}$,  $Q_{\mu\nu\lambda}=Q_{\nu\mu\lambda}$.

From a dynamical point of view, the metric and the affine connection are to be considered independent variables; we call metric-affine gravity any theory of spacetime that is constructed with such fundamental prescription. When comparing MAG and GR, the most natural action to consider is the generalization of the Einstein-Hilbert action, given by the only nontrivial full contraction of the Riemann tensor \eref{eq:riemannDef}, namely

\begin{eqnarray}
\label{eq:einsteinMag}
S[g, \Gamma] &= \int {\rm d}^4 x\sqrt{g} \; R^{\mu\nu}{}_{\mu\nu}\\
	&= \int {\rm d}^4 x\sqrt{g}\,\left[ {\mathring{R}} + \mathring{\nabla}_\mu \Phi^{\mu\lambda}{}_\lambda - \mathring{\nabla}_\lambda \Phi^{\mu\lambda}{}_\mu  + \Phi^\mu{}_{\gamma\mu} \Phi^{\gamma\nu}{}_\nu - \Phi_{\mu\gamma\nu} \Phi^{\gamma\nu\mu}
		\right]\,,
\end{eqnarray}
which is quadratic in torsion and non-metricity and where $\mathring{R}$ is the Ricci scalar computed from the Christoffel symbols of the background metric $g$. One can easily check that the equations of motion for the distortion obtained by Eq.~\eref{eq:einsteinMag} impose vanishing torsion and non-metricity \cite{Palatini:1919,Dadhich:2012htv,Iosifidis:2021tvx}. Hence, the MAG version of the Einstein-Hilbert action turns out to be dynamically equivalent to GR. This statement can actually be generalized to all the possible MAG actions that are quadratic in torsion and non-metricity with a non-degenerate quadratic form \cite{Iosifidis:2021tvx}, as long as we take into account only dimension-$2$ Lagrangians. From an effective field theory perspective, this statement is crucially dependent on the mass dimension of the operators $\sim T^2$ and $\sim Q^2$ and, therefore, only holds in the IR. This further motivates the development of the perturbation theory formalism for MAG degrees of freedom.

Upon inclusion of matter fields, we notice that torsion couples naturally to all spinors \cite{Gasperini:2017ggf,Wheeler:2023qyy}, while non-metricity does not. Moreover, a spin-parity analysis in flat space shows that non-metricity generally includes a propagating spin-$3$ particle \cite{Baldazzi:2021kaf,2647462}. For these reasons, from now on we will focus on torsion only. Since we will assume a Riemannian background, we will drop the small circle on top of the Christoffel symbols and the associated covariant derivatives and curvature tensors in the rest of this paper.

\section{Transverse decomposition of the torsion}\label{sect:Transv}

In this section, we want to generalize the transverse decomposition of York \cite{York:1974psa} to the torsion tensor. To this end, let us take into account the non-exhaustive algebraic decomposition of the torsion
\begin{eqnarray}
 T^\rho{}_{\mu\nu} = t^\rho{}_{\mu\nu} + H^\rho{}_{\mu\nu} \, ,
\end{eqnarray}
where $t$ is hook antisymmetric and $H$ is completely antisymmetric. The former tensor bears the information of vector and purely tensor torsion, while the latter is related to the axial torsion. We remind the reader that the properties of a hook antisymmetric tensor are
\begin{eqnarray}
 t^\rho{}_{[\mu\nu]} = t^\rho{}_{\mu\nu} \, , \qquad t_{\rho\mu\nu} + t_{\mu\nu\rho} + t_{\nu\rho\mu} = 0 \, ,
\end{eqnarray}
where the cyclicity property derives from the requirement that $t$ has a completely antisymmetric part equal to zero.

York defines the transverse decomposition of a symmetric rank-$(0,2)$ tensor $\phi$, such as the inverse of the metric, as
\begin{eqnarray}
 \phi^{\mu\nu} = \phi_{\tiny \perp}{}^{\mu\nu} + (K \xi)^{\mu\nu} \, ,
\end{eqnarray}
where $K$ is the Killing form of the vector $\xi^\mu$, $(K \xi)^{\mu\nu} = \nabla^\mu \xi^\nu + \nabla^\nu \xi^\mu$. Since $\phi^{\mu\nu}$ is a symmetric tensor, once we single out the transverse part, only vectors like $\xi^\mu$ can contribute to the one-time longitudinal modes of the field. This is expected also from the spin-parity decomposition of symmetric tensors, where the transverse part has $2^{+}$ and $0^{+}$ modes, while the longitudinal one has a $1^{-} $ and one additional $0^{+}$ \cite{Percacci:2017fkn}. In the following, we mimic the rationale at the base of York's work and apply it to the modes appearing in the torsion tensor.

\subsection{Transverse decomposition of the hook antisymmetric modes}

When we deal with hook antisymmetric rank-$(2,1)$ tensors like $t^{\rho}{}_{\mu\nu}$, the one-time longitudinal modes are parameterized by a symmetric tensor $S^{\mu\nu}$ and an antisymmetric one $A_{\mu\nu}$ \cite{Baldazzi:2021kaf}. Therefore, by analogy with \cite{York:1974psa}, we propose the transverse decomposition
\begin{eqnarray}
 t^\rho{}_{\mu\nu} = t_{\tiny \perp}{}^\rho{}_{\mu\nu} + (L_1 S)^\rho{}_{\mu\nu} + (L_2 A)^\rho{}_{\mu\nu} \, ,
\end{eqnarray}
where the generalizations of the Killing form read
 \begin{eqnarray}\label{L_1 S}
  & (L_1 S)^\rho{}_{\mu\nu} = \nabla_\mu S^\rho{}_\nu - \nabla_\nu S^\rho{}_\mu \, ;\\\label{L_1 A}
  & (L_2 A)^\rho{}_{\mu\nu} = 2 \nabla^\rho A_{\mu\nu} + \nabla_\mu A^\rho{}_\nu - \nabla_\nu A^\rho{}_\mu \, .
 \end{eqnarray}
It is easy to show that the cycle of these two tensor structures vanishes identically for torsion-free background connections. Notice that invariances with the same tensor structure of these differential operators were proposed in \cite{Curtright:1980yk} for hook antisymmetric tensors.

Let us take into account the two $GL(d)$-independent divergences of the hook antisymmetric part of the torsion. These are given by an antisymmetric tensor and a symmetric one, and a convenient basis is given by
\begin{eqnarray}\label{div12-t}
&(div_1 t)_{\mu\nu} \equiv \nabla_\rho t^\rho{}_{\mu\nu} \equiv (\Delta_1 S)_{\mu\nu} + (\Delta_2 A)_{\mu\nu} \, ,\\
&(div_2 t)^{\mu\nu} \equiv \nabla_\rho t^{\mu\rho\nu} + \nabla_\rho t^{\nu\rho\mu} \equiv (\Delta_3 S)^{\mu\nu} + (\Delta_4 A)^{\mu\nu} \, .
\end{eqnarray}
The choice of the basis is justified by the following identity
\begin{eqnarray}
 \nabla_\rho t^{\mu\rho\nu} - \nabla_\rho t^{\nu\rho\mu} = \nabla_\rho t^{\rho\mu\nu} \, .
\end{eqnarray}
On the right-hand sides of \eref{div12-t} we have implicitly defined the quadratic differential operators acting on the two tensors of our decomposition. {Decomposing $S^{\mu\nu}$ into $SO(d)$ irreducible components as $S^{\mu\nu}=\overline{S}{}^{\mu\nu}+\frac{1}{d} g^{\mu\nu} s$, the explicit form of the first divergence is}
\begin{eqnarray}\label{eq:div1-T}
(div_1 t)_{\mu\nu} = &  \nabla_\mu \nabla_\rho \overline{S}{}^\rho{}_\nu - \nabla_\nu \nabla_\rho \overline{S}{}^\rho{}_\mu + R_{\rho\mu} \overline{S}{}^\rho{}_\nu - R_{\rho\nu} \overline{S}{}^\rho{}_\mu\\\nonumber
& + 2 \square A_{\mu\nu} + \nabla_\mu \nabla_\rho A^\rho{}_\nu - \nabla_\nu \nabla_\rho A^\rho{}_\mu + R_{\rho\mu} A^\rho{}_\nu - R_{\rho\nu} A^\rho{}_\mu - R^{\rho\lambda}{}_{\mu\nu} A_{\rho\lambda} \, ,
\end{eqnarray}
while for the second one we find
\begin{eqnarray}
(div_2 t)^{\mu\nu} = & 2 \square \overline{S}{}^{\mu\nu} - \nabla^\mu \nabla_\rho \overline{S}{}^{\rho\nu} - \nabla^\nu \nabla_\rho \overline{S}{}^{\rho\mu} + \frac{2}{d} g^{\mu\nu} \square s - \frac{2}{d} \nabla^\mu \nabla^\nu s \\\nonumber
& + 2 R^\mu{}_\lambda{}^\nu{}_\rho \overline{S}{}^{\lambda\rho} - R_\rho{}^\mu \overline{S}{}^{\nu\rho} - R_\rho{}^\nu \overline{S}{}^{\mu\rho}\\
& + 3 \left( \nabla^\mu \nabla_\rho A^{\rho\nu} + \nabla^\nu \nabla_\rho A^{\rho\mu} + R^\mu{}_\rho A^{\rho\nu} + R^\nu{}_\rho A^{\rho\mu} \right) \, .\nonumber
\end{eqnarray}
Notice that, since $t^\rho{}_{\mu\nu}$ need not be trace-free, the second divergence has both trace-free and pure trace parts
\begin{eqnarray}
&(\overline{div_2 t})^{\mu\nu} \equiv ({div_2 t})^{\mu\nu} - \frac{1}{d} g^{\mu\nu} (div_2 t)^\lambda{}_\lambda \equiv (\Delta_5 S)^{\mu\nu} + (\Delta_4 A)^{\mu\nu} \, , \\
&(div_2 t)^\mu{}_\mu \equiv (\Delta_6 S) \, ,
\end{eqnarray}
where the latter differential operator maps symmetric tensors to scalars. Notice that, since $(div_2 t)^\lambda{}_\lambda$ is independent of $A_{\mu\nu}$, we do not need to define the trace-free counterpart of $\Delta_4$.

{Each of the three $SO(d)$ irrep's ($A_{\mu\nu}$, $\overline{S}_{\mu\nu}$ and $s$) has to be determined by inverting the equation of the corresponding source.
The explicit form of the trace of $(div_2 t)^{\mu\nu}$ is}
\begin{eqnarray}\label{eq:div2-trace-T}
 (div_2 t)^\mu{}_\mu = - 2 \nabla_\mu \nabla_\nu \overline{S}{}^{\nu\mu} + \frac{2(d-1)}{d} \square s \, .
\end{eqnarray}
From this equation we easily find the trace-free part of the same divergence
\begin{eqnarray}\label{eq:div2-tracefree-T}
 (\overline{div_2 t})^{\mu\nu} = & \, 2 \square \overline{S}{}^{\mu\nu} - \nabla^\mu \nabla_\rho \overline{S}{}^{\rho\nu} - \nabla^\nu \nabla_\rho \overline{S}{}^{\rho\mu} + \frac{2}{d} g^{\mu\nu} \nabla_\lambda \nabla_\rho \overline{S}{}^{\rho\lambda}\\\nonumber
 & + \frac{2}{d^2} g^{\mu\nu} \square s - \frac{2}{d} \nabla^\mu \nabla^\nu s 
  + 2 R^\mu{}_\lambda{}^\nu{}_\rho \overline{S}{}^{\lambda\rho} - R_\rho{}^\mu \overline{S}{}^{\nu\rho} - R_\rho{}^\nu \overline{S}{}^{\mu\rho} \\\nonumber
 &+ 3 \left( \nabla^\mu \nabla_\rho A^{\rho\nu} + \nabla^\nu \nabla_\rho A^{\rho\mu} + R^\mu{}_\rho A^{\rho\nu} + R^\nu{}_\rho A^{\rho\mu} \right) \, .
\end{eqnarray}
From Eq.s\ \eref{eq:div1-T}, \eref{eq:div2-trace-T} and \eref{eq:div2-tracefree-T} we immediately read off the form of the $\Delta_i$ second-order differential operators, which need to be invertible in order for the decomposition to be unique
 \begin{eqnarray}
  (\Delta_2 A)_{\mu\nu}  &= 2 \square A_{\mu\nu} + \nabla_\mu \nabla_\rho A^\rho{}_\nu - \nabla_\nu \nabla_\rho A^\rho{}_\mu + R_{\rho\mu} A^\rho{}_\nu - R_{\rho\nu} A^\rho{}_\mu - R^{\rho\lambda}{}_{\mu\nu} A_{\rho\lambda} \, , \\
  (\Delta_5 S)^{\mu\nu}  &= 2 \square \overline{S}{}^{\mu\nu} - \nabla^\mu \nabla_\rho \overline{S}{}^{\rho\nu} - \nabla^\nu \nabla_\rho \overline{S}{}^{\rho\mu} + \frac{2}{d} g^{\mu\nu} \nabla_\lambda \nabla_\rho \overline{S}{}^{\rho\lambda} \\\nonumber
  &\quad + \frac{2}{d^2} g^{\mu\nu} \square s - \frac{2}{d} \nabla^\mu \nabla^\nu s 
   + 2 R^\mu{}_\lambda{}^\nu{}_\rho \overline{S}{}^{\lambda\rho} - R_\rho{}^\mu \overline{S}{}^{\nu\rho} - R_\rho{}^\nu \overline{S}{}^{\mu\rho} \, , \\
  (\Delta_6 S)  &= - 2 \nabla_\mu \nabla_\nu \overline{S}{}^{\nu\mu} + \frac{2(d-1)}{d} \square s \, .
 \end{eqnarray}

We now verify the existence and uniqueness of the tensors $A_{\mu\nu}$ and $S_{\mu\nu}$, i.e., we show the existence and uniqueness of the decomposition itself. To this end, we check the Hermiticity and ellipticity of $\Delta_i$, $i=2,5,6$. The Hermiticity is easily verified by integrating by parts and using the algebraic symmetries of the Riemann tensor, $A_{\mu\nu}$ and $\overline{S}{}^{\mu\nu}$. Given generic tensors $B_{\mu\nu} = B_{[\mu\nu]}$ and $C_{\mu\nu}=C_{(\mu\nu)}$, we find
\begin{eqnarray}
 \int \sqrt{g} B^{\mu\nu} (\Delta_2 A)_{\mu\nu} = \int \sqrt{g} A^{\mu\nu} (\Delta_2 B)_{\mu\nu} \, ,
\end{eqnarray}
and
\begin{eqnarray}
 &\int \sqrt{g} (\overline{C}_{\mu\nu} + \frac{c}{d} g_{\mu\nu}) (\Delta_5 S)^{\mu\nu} = \int \sqrt{g} (\overline{S}_{\mu\nu} + \frac{s}{d} g_{\mu\nu}) (\Delta_5 C)^{\mu\nu} \, ,\\
& \int \sqrt{g} \, c \, (\Delta_6 s) = \int \sqrt{g} \, s \, (\Delta_6 c) \, .\nonumber
\end{eqnarray}

At this point, we are interested in inverting $\Delta_2$, $\Delta_5$ and $\Delta_6$ to show that, for some given $(div_1 t)_{\mu\nu}$, $(\overline{div_2 t}){}^{\mu\nu}$ and $(div_2 t)^\lambda{}_\lambda$, there exist unique tensors $A_{\mu\nu}$, $\overline{S}_{\mu\nu}$ and $s$ up to zero modes. For these operators to be invertible, we need the principal symbols $\boldsymbol{\sigma}_p (\Delta_i)$, $i=2, 5, 6$ to be of maximal rank. We only take into account the action of $\Delta_3$ on the trace-free part of $S_{\mu\nu}$, for the trace mode is to be determined by inverting $\Delta_5$. Therefore, we start by writing the part of $\Delta_3$ which acts on the symmetric traceless tensor $\overline{S}{}^{\mu\nu}$
\begin{eqnarray}
 \left[(\boldsymbol{\sigma}_p (\Delta_5) ) \overline{S}\right]^{\mu\nu} = 2 p^2 \overline{S}{}^{\mu\nu} - p^\mu p_\rho \overline{S}{}^{\rho\nu} - p^\nu p_\rho \overline{S}{}^{\rho\mu} + 2 g^{\mu\nu} p_\rho p_\lambda \overline{S}{}^{\rho\lambda}  \, .
\end{eqnarray}
We decompose $\overline{S}{}^{\mu\nu}$ in its transverse and longitudinal components. Some words of caution are now in order. Usually, a symmetric traceless rank-$2$ tensor has a spin-$0$ part which is twice longitudinal. However, such a component cannot appear here since it would violate the spin-parity decomposition of the torsion in the flat space limit \cite{Sezgin:1981xs,Baldazzi:2021kaf}. Therefore, in this special case, we have
\begin{eqnarray}\label{S-decomp-1}
 \overline{S}{}^{\mu\nu} = {\overline{S}_{\tiny \perp}}{}^{\mu\nu} + i p^\mu {{\zeta}_{\tiny \perp}}{}^\nu + i p^\nu {{\zeta}_{\tiny \perp}}{}^\mu \, ,
\end{eqnarray}
where both ${\overline{S}_{\tiny \perp}}{}^{\mu\nu}$ and $\zeta_{\tiny \perp}{}^\mu$ are transverse. Plugging such decomposition into the principal symbol of $\Delta_5$ we find
\begin{eqnarray}
\label{eq:princsymbol}
 \left[(\boldsymbol{\sigma}_p (\Delta_5) ) \overline{S}\right]^{\mu\nu} = 2 p^2 {\overline{S}_{\tiny \perp}}{}^{\mu\nu} + i p^2 \left( p^\mu {{\zeta}_{\tiny \perp}}{}^\nu + p^\nu {{\zeta}_{\tiny \perp}}{}^\mu \right) \, .
\end{eqnarray}
The ellipticity of the $\overline{S}_{\tiny \perp}^{\mu\nu}$ sector is obvious from the above expression. Concerning the subspace spanned by $\zeta_{\perp}^\mu$, one could act with a further derivative on the l.h.s. of Eq.\ \eref{eq:princsymbol} to project onto the longitudinal component of $\overline{S}{}^{\mu\nu}$ and verify that $\Delta_5$ is indeed elliptic.

Let us now to $A_{\mu\nu}$. In this case, we obtain the following principal symbol for $\Delta_2$ 
\begin{eqnarray}
 \left[(\boldsymbol{\sigma}_p (\Delta_2) ) A\right]{}_{\mu\nu} = 2 p^2 A_{\mu\nu} + p_\mu p_\rho A^\rho{}_\nu - p_\nu p_\rho A^\rho{}_\mu \, .
\end{eqnarray}
Exactly as before, we proceed by decomposing $A_{\mu\nu}$ into its transverse part ${A_{\tiny \perp}}{}_{\mu\nu}$ and its longitudinal one, which is parametrized by the transverse vector $\xi_\mu$
\begin{eqnarray}\label{A-decomp-1}
 A_{\mu\nu} = {A_{\tiny \perp}}{}_{\mu\nu} + i p_\mu {{\xi}_{\tiny \perp}}{}_\nu - i p_\nu {{\xi}_{\tiny \perp}}{}_\mu \, .
\end{eqnarray}
Inserting this decomposition into the expression of the principal symbol we obtain
\begin{eqnarray}
 \left[(\boldsymbol{\sigma}_p (\Delta_2) ) A\right]{}_{\mu\nu} = 2 p^2 {A_{\tiny \perp}}{}_{\mu\nu} + i p^2 \left( p_\mu {{\xi}_{\tiny \perp}}{}_\nu - p_\nu {{\xi}_{\tiny \perp}}{}_\mu \right) \, .
\end{eqnarray}
Given the minus sign in the terms of the above equation involving ${{\xi}_{\tiny \perp}}{}_\mu$, one may worry that some nontrivial configuration might reduce the rank of $\boldsymbol{\sigma}_p (\Delta_2)$ in the vector sector. However, since ${{\xi}_{\tiny \perp}}{}_\mu$ is transverse from the onset, we notice that also this operator is elliptic.

At last, the ellipticity of $\Delta_5$ acting on the trace modes is easily seen from its principal symbol
\begin{eqnarray}
 ((\boldsymbol{\sigma}_p(\Delta_6))S) =  \frac{2(d-1)}{d} p^2 s \, ,
\end{eqnarray}
where we have exploited the further decomposition of the traceless part of $S^{\mu\nu}$.

At this point, the only trouble which may arise when we are about to invert these operators to solve for $S^{\mu\nu}$ and $A_{\mu\nu}$ is the presence of zero modes. First of all, notice that, since
 \begin{eqnarray}
  & \int \sqrt{g} (L_1 S)^\rho{}_{\mu\nu} (L_1 S)_\rho{}^{\mu\nu} = - \int \sqrt{g} S_{\mu\nu} (div_2 (L_1 S) )^{\mu\nu} \, , \label{eq:(L_1)^2}\\
  & \int \sqrt{g} (L_2 A)^\rho{}_{\mu\nu} (L_2 A)_\rho{}^{\mu\nu} = - 3 \int \sqrt{g} A_{\mu\nu} (div_1 (L_2 A))^{\mu\nu} \, ,\label{eq:(L_2)^2}
 \end{eqnarray}
the zero modes of $\Delta_i$, $i=2,5,6$ are exactly the zero modes of the Curtright forms \eref{L_1 S} and \eref{L_1 A}.
However, these modes do not invalidate the inversion of the operators on closed manifolds as long as they are globally orthogonal to their sources \cite{York:1974psa}. Indeed, this is the case for both the $GL(d)$-independent sources
 \begin{eqnarray}  
  & \int \sqrt{g} S_{\rho\nu} (div_2 t)^{\rho\nu} = - \int \sqrt{g} (L_1 S)^\rho{}_{\mu\nu} t_\rho{}^{\mu\nu} \, , \\
  & \int \sqrt{g} A_{\mu\nu} (div_1 t)^{\mu\nu} = - \frac{1}{3} \int \sqrt{g} (L_2 A)^\rho{}_{\mu\nu} t_\rho{}^{\mu\nu} \, ,
 \end{eqnarray}
for $(L_1 S)^\rho{}_{\mu\nu} = 0$ and $(L_2 A)^\rho{}_{\mu\nu} = 0$ are the definitions of the zero-modes. In passing, we note that these two equations are just the general case of the two previous ones \eref{eq:(L_1)^2} and \eref{eq:(L_2)^2}.

\subsection{Transverse decomposition of the {axial modes}}
\label{Transv-TA}

After having studied in detail the transverse decomposition of the hook antisymmetric part of the torsion $t^\rho{}_{\mu\nu}$, we focus now on the completely antisymmetric component $H^\rho{}_{\mu\nu}$. Exploiting the Hodge duality, the latter tensor can be traded for its dual rank-$(d-3)$ completely antisymmetric tensor as
\begin{eqnarray}
\theta_{\mu_1\dots \mu_{d-3}} \equiv \varepsilon_{\mu_1\dots\mu_{d-3}\lambda}{}^{\alpha\beta} T^\lambda{}_{\alpha\beta} = \varepsilon_{\mu_1\dots\mu_{d-3}\lambda}{}^{\alpha\beta} H^\lambda{}_{\alpha\beta} \, .
\end{eqnarray}
For the rest of the paper, we are going to denote the axial component through its Hodge dual $\theta$. 
Note that only in $d=4$ both $H$ and $\theta$ naturally have the same Weyl weight, which of course is zero since these tensors are components of an affine connection.

The spin-parity content of completely antisymmetric tensors is well-known \cite{Baldazzi:2021kaf} and we readily find the following covariant decomposition
\begin{eqnarray}\label{eq:H-decomp}
 H^\rho{}_{\mu\nu} = \frac{1}{3!(d-3)!} \varepsilon^{\sigma_1\dots\sigma_{d-3}\rho}{}_{\mu\nu} \left( {\theta_{\tiny \perp}}_{\sigma_1\dots\sigma_{d-3}} + \nabla_{[\sigma_1} {\pi_{\tiny \perp}}_{\sigma_2\dots\sigma_{d-3}]} \right) \, .
\end{eqnarray}
Notice that this result is in complete agreement with the Hodge decomposition of $p$-forms \cite{Hodge:1989}, the only difference being that we have not singled out the harmonic part of $\theta$.

In the next section, we discuss a decomposition of the full torsion $2$-form which is both transverse and traceless. Since the completely antisymmetric part has (trivially) no trace modes, we shall be able to exploit the result of Eq.\ \eref{eq:H-decomp} also in the next section.

\section{Transverse-traceless decomposition}\label{sect:TransvT}

We now elaborate on a decomposition that is both transverse and traceless. As noticed above, we can safely focus on the hook antisymmetric part of $T^\rho{}_{\mu\nu}$ only. Firstly, we check that the trace of the transverse part ${t_{\perp}}^\rho{}_{\mu\nu}$ is transverse
\begin{eqnarray}
\nabla^\mu {t_{\tiny \perp}}^\rho{}_{\mu\rho} = \delta^\nu{}_\rho \nabla^\mu {t_{\tiny \perp}}^\rho{}_{\mu\nu} = 0 \, .
\end{eqnarray}
However, the trace-free part of the transverse piece is not transverse under the first divergence
\begin{eqnarray}
\nabla_\rho \left[ {t_{\tiny \perp}}^{\rho}{}_{\mu\nu} - \frac{1}{d-1} \left(\delta^\rho{}_\nu {t_{\tiny \perp}}^\lambda{}_{\mu\lambda} - \delta^\rho{}_\mu {t_{\tiny \perp}}^\lambda{}_{\nu\lambda}\right) \right] = - \frac{1}{d-1} \left( \nabla_\nu {t_{\tiny \perp}}^\rho{}_{\mu\rho} - \nabla_\mu {t_{\tiny \perp}}^\rho{}_{\nu\rho} \right) \neq 0 \, .
\end{eqnarray}
Analogously, the same object is not transverse in its last two indices as well
\begin{eqnarray}
\nabla^\mu \left[ {t_{\tiny \perp}}^{\rho}{}_{\mu\nu} - \frac{1}{d-1} \left(\delta^\rho{}_\nu {t_{\tiny \perp}}^\lambda{}_{\mu\lambda} - \delta^\rho{}_\mu {t_{\tiny \perp}}^\lambda{}_{\nu\lambda}\right) \right] = \frac{1}{d-1} \nabla^\rho {t_{\tiny \perp}}^\lambda{}_{\nu\lambda} \neq 0 \, .
\end{eqnarray}
Therefore, we need to build a new decomposition that is both transverse and traceless. To this end, let us first divide the hook antisymmetric part of the torsion into its purely tensorial and vector components
\begin{eqnarray}\label{eq:traceless-decomp-t}
 t^\rho{}_{\mu\nu} = \kappa^\rho{}_{\mu\nu} + \frac{1}{d-1} \left( \delta^\rho{}_\nu \tau_\mu - \delta^\rho{}_\mu \tau_\nu \right) \, ,
\end{eqnarray}
where $\tau_\mu \equiv t^\rho{}_{\mu\rho}$. There is a striking difference between this tracefree and pure trace algebraic decomposition and the analog for the metric perturbations: the former is $GL(d)$-invariant, while the latter is only $SO(d)$-invariant. In analogy with the transverse decomposition, we write
\begin{eqnarray}\label{TT-decomposition}
 \kappa^\rho{}_{\mu\nu} = {\kappa_{\tiny \perp}}^\rho{}_{\mu\nu} + (M_1 S)^\rho{}_{\mu\nu} + (M_2 A)^\rho{}_{\mu\nu} \, ,
\end{eqnarray}
where $\kappa_{\tiny \perp}$ is transverse traceless and tracefree Curtright forms are
 \begin{eqnarray}
   (M_1 S)^\rho{}_{\mu\nu} = & \nabla_\mu S^\rho{}_\nu - \nabla_\nu S^\rho{}_\mu\\
  &\, - \frac{1}{d-1} \left( \delta^\rho{}_\nu (\nabla_\mu s - \nabla_\lambda S^\lambda{}_\mu) - \delta^\rho{}_\mu (\nabla_\nu s - \nabla_\lambda S^\lambda{}_\nu) \right) \, ;\nonumber\\\label{eq:M_2 A}
   (M_2 A)^\rho{}_{\mu\nu} = & 2 \nabla^\rho A_{\mu\nu} + \nabla_\mu A^\rho{}_\nu - \nabla_\nu A^\rho{}_\mu - \frac{3}{d-1} \left( \delta^\rho{}_\nu \nabla^\lambda A_{\mu\lambda} - \delta^\rho{}_\mu \nabla^\lambda A_{\nu\lambda} \right) \, .
 \end{eqnarray}
The $M_i$'s operators differ from the $L_i$'s in that they consistently give no contribution to the trace. While the tracefree Killing form which appears in the transverse traceless decomposition of symmetric $2$-tensors has simple conformal properties, no such feature arises in the present case. Let us notice that the trace of $S$ automatically decouples from the previous expression, so we can identically substitute $S=\overline{S}$ into the first equation, which becomes
\begin{eqnarray}\label{eq:M_1 S}
 (M_1 S)^\rho{}_{\mu\nu} = \nabla_\mu \overline{S}{}^\rho{}_\nu - \nabla_\nu \overline{S}{}^\rho{}_\mu + \frac{1}{d-1} \left( \delta^\rho{}_\nu \nabla_\lambda \overline{S}{}^\lambda{}_\mu - \delta^\rho{}_\mu \nabla_\lambda \overline{S}{}^\lambda{}_\nu \right) \, .
\end{eqnarray}

We proceed now with the same reasoning exploited in the previous section, and first compute the two independent divergences of $\kappa$. Obviously, in this case we need not to single out the trace of the second divergence, for $\kappa$ is already traceless. The divergences are defined exactly as before, and their explicit expressions are given by
\begin{eqnarray}
 (div_1 \kappa)_{\mu\nu} = & \, \frac{d-2}{d-1} \left( \nabla_\mu \nabla_\rho \overline{S}{}^\rho{}_\nu - \nabla_\nu \nabla_\rho \overline{S}{}^\rho{}_\mu \right)  + R_{\rho\mu} \overline{S}{}^\rho{}_\nu - R_{\rho\nu} \overline{S}{}^\rho{}_\mu \\\nonumber
 & + 2 \square A_{\mu\nu} + \frac{d+2}{d-1} \left( \nabla_\mu \nabla_\rho A^\rho{}_\nu - \nabla_\nu \nabla_\rho A^\rho{}_\mu \right)\nonumber\\
& + R_{\rho\mu} A^\rho{}_\nu - R_{\rho\nu} A^\rho{}_\mu - R^{\rho\lambda}{}_{\mu\nu} A_{\rho\lambda} \, , \nonumber\\
 (div_2 \kappa)^{\mu\nu} = & \, 2 \square \overline{S}{}^{\mu\nu} - \frac{d}{d-1} \nabla^\mu \nabla_\rho \overline{S}{}^{\rho\nu} - \frac{d}{d-1} \nabla^\nu \nabla_\rho \overline{S}{}^{\rho\mu} +\frac{2}{d-1} g^{\mu\nu} \nabla_\rho \nabla_\lambda \overline{S}{}^{\lambda\rho}\\
 & - R_\rho{}^\mu \overline{S}{}^{\nu\rho} - R_\rho{}^\nu \overline{S}{}^{\mu\rho} + 2 R^\mu{}_\lambda{}^\nu{}_\rho \overline{S}{}^{\lambda\rho}\nonumber\\
 & + \frac{3(d-2)}{d-1} \left( \nabla^\mu \nabla_\rho A^{\rho\nu} + \nabla^\nu \nabla_\rho A^{\rho\mu} \right) + 3 \left( R^\mu{}_\rho A^{\rho\nu} + R^\nu{}_\rho A^{\rho\mu} \right) \, .\nonumber
\end{eqnarray}
Notice that the previous expressions differ from the ones found in the transverse decomposition only in the coefficients of the non-minimal derivative terms. As a consistency check, we observe that the trace of the second equation vanishes identically.

In complete analogy with the procedure of the previous section, we can single out the second-order differential operators acting on $\overline{S}_{\mu\nu}$ and $A_{\mu\nu}$ from the two independent divergences of $\kappa$. The defining relations are
 \begin{eqnarray}
 & (div_1 t)_{\mu\nu} \equiv (\overline{\Delta}_1 \overline{S})_{\mu\nu} + (\overline{\Delta}_2 A)_{\mu\nu} \, , \\
 & (div_2 t)^{\mu\nu} \equiv (\overline{\Delta}_3 \overline{S})^{\mu\nu} + (\overline{\Delta}_4 A)^{\mu\nu} \, .
 \end{eqnarray}
By comparison with the expressions of the two divergences we find the explicit form of the two operators which display Laplacians of the $\overline{S}$ and $A$ tensors
 \begin{eqnarray}
   (\overline{\Delta}_2 A)_{\mu\nu} &= 2 \square A_{\mu\nu} + \frac{d+2}{d-1} \left( \nabla_\mu \nabla_\rho A^\rho{}_\nu - \nabla_\nu \nabla_\rho A^\rho{}_\mu \right)\\
  &\quad + R_{\rho\mu} A^\rho{}_\nu - R_{\rho\nu} A^\rho{}_\mu - R^{\rho\lambda}{}_{\mu\nu} A_{\rho\lambda} \, , \nonumber\\
  (\overline{\Delta}_3 S)^{\mu\nu} &= 2 \square \overline{S}{}^{\mu\nu} - \frac{d}{d-1} \nabla^\mu \nabla_\rho \overline{S}{}^{\rho\nu} - \frac{d}{d-1} \nabla^\nu \nabla_\rho \overline{S}{}^{\rho\mu} +\frac{2}{d-1} g^{\mu\nu} \nabla_\rho \nabla_\lambda \overline{S}{}^{\lambda\rho}\\\nonumber
  & \quad - R_\rho{}^\mu \overline{S}{}^{\nu\rho} - R_\rho{}^\nu \overline{S}{}^{\mu\rho} + 2 R^\mu{}_\lambda{}^\nu{}_\rho \overline{S}{}^{\lambda\rho} \, .
 \end{eqnarray}
As already observed, only the coefficients of the non-minimal parts of the operators acting on $\overline{S}_{\mu\nu}$ and $A_{\mu\nu}$ have changed. Since the hermicity of the operators of the previous section does not depend on the precise value of these constants, the operators are still hermitian, i.e.,
\begin{eqnarray}
 &\int \sqrt{g} B^{\mu\nu} \left( \overline{\Delta}_2 A \right){}_{\mu\nu} = \int \sqrt{g} A^{\mu\nu} \left( \overline{\Delta}_2 B \right){}_{\mu\nu} \, , \\
 &\int \sqrt{g} \, \overline{C}^{\mu\nu} \left( \overline{\Delta}_3 \overline{S} \right){}_{\mu\nu} = \int \sqrt{g} \, \overline{S}{}^{\mu\nu} \left( \overline{\Delta}_3 \overline{C} \right){}_{\mu\nu} \, ,
\end{eqnarray}
where we have exploited the fact that $(\overline{\Delta}_3 C)=0$ if $C^{\mu\nu}= \frac{1}{d} g^{\mu\nu} C^\lambda{}_\lambda$, i.e., the trace modes are in the kernel of $\overline{\Delta}_3$.
Exactly as in the transverse decomposition, we are now interested in checking the ellipticity of the operators $\overline{\Delta}_3$ and $\overline{\Delta}_2$. Making use of the same spin-parity decomposition used in the previous section for $S_{\mu\nu}$ and $A_{\mu\nu}$, we find the following principal symbols
 \begin{eqnarray}
 & \left[(\boldsymbol{\sigma}_p(\overline{\Delta}_3)) \overline{S}\right]^{\mu\nu} = 2 p^2 \overline{S}_{\tiny \perp}{}^{\mu\nu} + \frac{ip^2(d-2)}{d-1} \left( p^\mu \zeta_{\tiny \perp}{}^\nu + p^\nu \zeta_{\tiny \perp}{}^\mu \right) \, ; \\
 & \left[(\boldsymbol{\sigma}_p(\overline{\Delta}_2)) A\right]_{\mu\nu} = 2 p^2 {A_{\tiny \perp}}{}_{\mu\nu} + \frac{3id p^2}{d-1} \left( p_\mu \xi_{\tiny \perp}{}_\nu - p_\nu \xi_{\tiny \perp}{}_\mu \right) \, .
 \end{eqnarray}
These symbols are non-degenerate in each spin-parity sector, exactly as in the preceding section, therefore the operators are elliptic.

At this point, we focus on the overall matter content of our decomposition. To this end, we compare the flat-space limit of the transverse-traceless decomposition we have just worked out with the known flat-space spin-parity decomposition of the hook antisymmetric tensor $t^\rho{}_{\mu\nu}$. The latter may be written as
\begin{eqnarray}\label{flat-space-TT}
 t^\rho{}_{\mu\nu} = & \, {\overline{t}}_{\tiny \perp}{}^\rho{}_{\mu\nu} + \frac{1}{d-1} \left( \delta^\rho{}_\nu \tau_{\tiny \perp}{}_\mu - \delta^\rho{}_\mu \tau_{\tiny \perp}{}_\nu \right) + \frac{1}{d-1} \left( \delta^\rho{}_\nu \partial_\mu \varphi - \delta^\rho{}_\mu \partial_\nu \varphi \right)\\
 & \, + \partial_\mu X_{\tiny \perp}{}^\rho{}_\nu - \partial_\nu X_{\tiny \perp}{}^\rho{}_\mu + 2 \partial^\rho Y_{\tiny \perp}{}_{\mu\nu} + \partial_\mu Y_{\tiny \perp}{}^\rho{}_\nu - \partial_\nu Y_{\tiny \perp}{}^\rho{}_\mu\nonumber\\
 &\, + \partial_\mu \partial^\rho V_{\tiny \perp}{}_\nu - \partial_\nu \partial^\rho V_{\tiny \perp}{}_\mu + \frac{1}{d-1} \left( \delta^\rho{}_\nu \square V_{\tiny \perp}{}_\mu - \delta^\rho{}_\mu \square V_{\tiny \perp}{}_\nu \right) \, ,\nonumber
\end{eqnarray}
where all the tensors appearing on the r.h.s.\ of the above equation are transverse. To make contact with this expression, we have to further covariantly decompose the $\overline{S}$ and $A$ tensors into their spin-parity components as
 \begin{eqnarray}\label{S-split}
  & \overline{S}{}^{\mu\nu} = {\overline{S}}_{\tiny \perp}{}^{\mu\nu} + \nabla^\mu \zeta_{\tiny \perp}{}^\nu + \nabla^\nu \zeta_{\tiny \perp}{}^\mu \, , \label{A-split}\\
  & A_{\mu\nu} = A_{\tiny \perp}{}_{\mu\nu} + \nabla_\mu \xi_{\tiny \perp}{}_\nu - \nabla_\nu \xi_{\tiny \perp}{}_\mu \, .
 \end{eqnarray}
Here all the tensors and vectors that appear on the right-hand sides of the two equations are transverse. In complete analogy, we also decompose the trace $\tau_\mu$ as $\tau_\mu = \tau_{\tiny \perp}{}_\mu + \partial_\mu \varphi$, where $\nabla_\mu \tau_{\tiny \perp}{}^\mu = 0$. We plug these decompositions into Eq. \eref{TT-decomposition} and find
\begin{eqnarray}\label{full-TT-decomp-1}
 t^\rho{}_{\mu\nu} = & \, \kappa_{\tiny \perp}{}^\rho{}_{\mu\nu} + \frac{1}{d-1} \left( \delta^\rho{}_\nu \tau_{\tiny \perp}{}_\mu - \delta^\rho{}_\nu \tau_{\tiny \perp}{}_\mu \right) + \frac{1}{d-1} \left( \delta^\rho{}_\nu \partial_\mu \varphi - \delta^\rho{}_\mu \partial_\nu \varphi \right) \\\nonumber
 & \, + \nabla_\mu \overline{S}_{\tiny \perp}{}^\rho{}_\nu - \nabla_\nu \overline{S}_{\tiny \perp}{}^\rho{}_\mu + \nabla^\rho \nabla_\mu \zeta_{\tiny \perp}{}_\nu - \nabla^\rho \nabla_\nu \zeta_{\tiny \perp}{}_\mu\\
 &+ \frac{1}{d-1} \left[ \delta^\rho{}_\nu \left( \square \zeta_{\tiny \perp}{}_\mu - R_{\lambda\mu} \zeta_{\tiny \perp}{}^\lambda \right) - \delta^\rho{}_\mu \left( \square \zeta_{\tiny \perp}{}_\nu - R_{\lambda\nu} \zeta_{\tiny \perp}{}^\lambda \right) \right] \nonumber\\
 & \, + 2 \nabla^\rho A_{\tiny \perp}{}_{\mu\nu} + \nabla_\mu A_{\tiny \perp}{}^\rho{}_\nu - \nabla_\nu A_{\tiny \perp}{}^\rho{}_\mu + 3 \nabla^\rho \nabla_\mu \xi_{\tiny \perp}{}_\nu - 3 \nabla^\rho \nabla_\nu \xi_{\tiny \perp}{}_\mu \nonumber\\
 &+ \frac{3}{d-1} \left[ \delta^\rho{}_\nu \left( \square \xi_{\tiny \perp}{}_\mu - R_{\lambda\mu} \xi_{\tiny \perp}{}^\lambda \right) - \delta^\rho{}_\mu \left( \square \xi_{\tiny \perp}{}_\nu - R_{\lambda\nu} \xi_{\tiny \perp}{}^\lambda \right) \right] \, .\nonumber
\end{eqnarray}
Notice that we have exploited the first Bianchi identity to sort those terms that involve the vectors $\xi$ and $\zeta$ with $\nabla^\rho$ on the left. Moreover, observe that these two vectors appear through the very same tensor structures. Therefore, by comparing the last equation with the flat-space result Eq.\ \eref{flat-space-TT} we find
\begin{eqnarray}
 X_{\mu\nu} = \overline{S}_{\tiny \perp}{}_{\mu\nu} \, , \qquad Y_{\mu\nu} = A_{\tiny \perp}{}_{\mu\nu} \, , \qquad V_\mu = \zeta_{\tiny \perp}{}_\mu + 3 \xi_{\tiny \perp}{}_\mu \, .
\end{eqnarray}
Hence, we realize that the decomposition Eq.~\eref{TT-decomposition} is invariant under the following simultaneous transformations of $S_{\mu\nu}$ and $A_{\mu\nu}$
\begin{eqnarray}
A_{\mu\nu} & \rightarrow A_{\mu\nu}  + \nabla_{\nu}\Xi_{\tiny \perp}{}_{\mu} - \nabla_{\mu}\Xi_{\tiny \perp}{}_{\nu}\,,\\
S_{\mu\nu} & \rightarrow S_{\mu\nu} +3 \nabla_{\mu}\Xi_{\tiny \perp}{}_{\nu} - 3\nabla_{\nu}\Xi_{\tiny \perp}{}_{\mu}\,,
\end{eqnarray}
with $\Xi_{\tiny \perp}{}_\mu$ a transverse vector field.

We now turn to the analysis of zero-modes. In complete analogy with the previous case we find
 \begin{eqnarray}
  \int \sqrt{g} (M_1 \overline{S})^\rho{}_{\mu\nu} (M_1 \overline{S})_\rho{}^{\mu\nu} = - \int \sqrt{g} \overline{S}_{\mu\nu} (div_2 (M_1 \overline{S}))^{\mu\nu} \, , \\
  \int \sqrt{g} (M_2 A)^\rho{}_{\mu\nu} (M_2 A)_\rho{}^{\mu\nu} = - 3 \int \sqrt{g} A_{\mu\nu} (div_1 (M_2 A))^{\mu\nu} \, .
 \end{eqnarray}
The presence of the new tensor structures on the right-hand sides of Eqs.\ \eref{eq:M_2 A} and \eref{eq:M_1 S} does not alter the result, since both $(M_1 \overline{S})$ and $(M_2 A)$ are traceless. The previous equations are again telling us that the zero modes of $\overline{\Delta}_{2,3}$ are also zero-modes of the tracefree Curtright forms {\eref{eq:M_2 A} and \eref{eq:M_1 S}}. Using the same manipulations, we also see that the zero-modes of $A_{\mu\nu}$ and $\overline{S}{}^{\mu\nu}$ are globally orthogonal to their sources, as in the transverse case. Thence, the hermitian operators $\overline{\Delta}_i$, $i=2,3$ are always invertible on closed manifolds \cite{York:1974psa}. It is also clear that the longitudinal components $(M_1 A)^\rho{}_{\mu\nu}$ and $(M_2 S)^\rho{}_{\mu\nu}$ are orthogonal to the transverse part of $\kappa^\rho{}_{\mu\nu}$
\begin{eqnarray}
 \int \sqrt{g} (M_2 A)^\rho{}_{\mu\nu} \kappa_{\tiny \perp}{}_\rho{}^{\mu\nu} = - 3 \int \sqrt{g} A_{\mu\nu} \nabla_\rho \kappa_{\tiny \perp}{}^{\rho\mu\nu} = 0 \, .
\end{eqnarray}
A completely analogous formula holds for the scalar product of $(M_1 \overline{S})^\rho{}_{\mu\nu}$. Finally, the part of the decomposition which is longitudinal in two indices and is parametrized by a vector is also orthogonal to the transverse-traceless part by construction, because it has been obtained, up to some irrelevant overall constant, by taking either $A_{\mu\nu} \rightarrow \nabla_\mu \xi_{\tiny \perp}{}_\nu - \nabla_\nu \xi_{\tiny \perp}{}_\mu$ or, equivalently, $\overline{S}_{\mu\nu} \rightarrow \nabla_\mu \xi_{\tiny \perp}{}_\nu + \nabla_\nu \xi_{\tiny \perp}{}_\mu$.

\subsection{The special case of maximally symmetric spaces}\label{sbsbsect:MaxSym}

In this subsection, we specialize to the phenomenologically interesting case of a maximally symmetric background. The reader may have a de-Sitter spacetime in mind. In the previous analysis we have either assumed the absence of zero-modes of the operators $\overline{\Delta}_{2,3}$, or checked that they are orthogonal to their sources, and thence do not invalidate the inversion of the operators as long as we deal with closed manifolds. Nevertheless, let us now take them into account in a realistic physical application.

The two algebraic independent divergences of $\kappa^\rho{}_{\mu\nu}$ take a remarkably simple form when we specialize to maximally symmetric spaces
 \begin{eqnarray}
   (div_1 \kappa)_{\mu\nu} =& 2 \square A_{\tiny \perp}{}_{\mu\nu} + \frac{2(d-2)}{d(d-1)} R A_{\tiny \perp}{}_{\mu\nu}\\
   &+ 2 \nabla_\mu \left( \square \xi_{\tiny \perp}{}_\nu + \frac{2d-5}{d(d-1)} R \xi_{\tiny \perp}{}_\nu \right) - 2 \nabla_\nu \left( \square \xi_{\tiny \perp}{}_\mu + \frac{2d-5}{d(d-1)} R \xi_{\tiny \perp}{}_\mu \right) \, , \nonumber\\
   (div_2 \kappa)_{\mu\nu} =& 2 \square \overline{S}_{\tiny \perp}{}_{\mu\nu} - \frac{2}{d-1} R \overline{S}_{\tiny \perp}{}_{\mu\nu}\\
   & + 2 \nabla_\mu \left( \square \zeta_{\tiny \perp}{}_\nu + \frac{1}{d(d-1)} R \zeta_{\tiny \perp}{}_\nu \right) + 2 \nabla_\nu \left( \square \zeta_{\tiny \perp}{}_\mu + \frac{1}{d(d-1)} R \zeta_{\tiny \perp}{}_\mu \right) \, ,\nonumber
 \end{eqnarray}
where we have split the $\overline{S}$ and $A$ tensors as in Eqs\ \eref{S-split} and \eref{A-split}. Notice that the dependence on $\overline{S}_{\mu\nu}$ and $A_{\mu\nu}$ has identically decoupled from the first and second divergences respectively, i.e., we now have a decoupled system of differential equations. Observe also that, differently from the case of metric fluctuations, the assumption of Einstein spaces is not enough for such simplification to take place.

A sufficient condition for the presence of zero modes of the first divergence, or equivalently of the second tracefree Curtright form \eref{eq:M_2 A}, is the following system of equations
 \begin{eqnarray}
  & \square A_{\tiny \perp}{}_{\mu\nu} + \frac{d-2}{d(d-1)} R A_{\tiny \perp}{}_{\mu\nu} = 0 \, , \\
  & \square \xi_{\tiny \perp}{}_\mu + \frac{(2d-5)}{d(d-1)} R \xi_{\tiny \perp}{}_\mu = v_\mu \, ,
 \end{eqnarray}
where $v_\mu$ is a closed $1$-form.
A very similar analysis shows that a sufficient condition for zero-modes of $(div_2 \kappa)$ is
 \begin{eqnarray}
  & \square \overline{S}_{\tiny \perp}{}_{\mu\nu} - \frac{1}{d-1} R \overline{S}_{\tiny \perp}{}_{\mu\nu} = 0 \, ,\\
  & \square \zeta_{\tiny \perp}{}_\mu + \frac{1}{d(d-1)} R \zeta_{\tiny \perp}{}_\mu = K_\mu \, ,
 \end{eqnarray}
where now $K_\mu$ is any one of the $\frac{d(d+1)}{2}$ Killing vectors. Notice that, at least for $d \neq 3$, it is not possible to combine the equations for the two vectors so that only the physical linear combination $3 \xi_{\tiny \perp}{}_\mu + \zeta_{\tiny \perp}{}_\mu$ appears. However, in the flat-space limit, the conditions for the appearance of zero-modes for the two divergences simplify considerably to the following form
 \begin{eqnarray}
  & \square A_{\mu\nu} = \square A_{\tiny \perp}{}_{\mu\nu} + \partial_\mu \xi_{\tiny \perp}{}_\nu - \partial_\nu \xi_{\tiny \perp}{}_\mu = 0 \, ,\\
  & \square \overline{S}_{\mu\nu} = \square \overline{S}_{\tiny \perp}{}_{\mu\nu} + \partial_\mu \zeta_{\tiny \perp}{}_\nu + \partial_\nu \zeta_{\tiny \perp}{}_\mu = 0 \, .
 \end{eqnarray}
Note that the two equations are expressed in terms of the non-transverse tensors $A_{\mu\nu}$ and $\overline{S}_{\mu\nu}$. Thus, the zero modes of the trace-free Curtright forms {\eref{eq:M_2 A} and \eref{eq:M_1 S}} appear in the flat-space limit as harmonic tensors.

In the general case, we invert the operators $\overline{\Delta}_{2,3}$ assuming that their kernels are empty to find uniquely $A_{\mu\nu}$ and $\overline{S}_{\mu\nu}$ from $(div_1 \kappa)$ and $(div_2 \kappa)$, respectively. The complete solutions for these tensors are found by adding the solutions of the homogeneous equations, i.e., the zero-modes. As we have just seen, in the special case of small torsion fluctuations in flat-space, these zero-modes are harmonic tensors, whose spectrum is well-known. In any case, since the kernels of $\Delta_{2,3}$ are isomorphic to those of the tracefree Curtright forms, the zero modes never affect the decomposition itself.

\subsection{The full transverse-traceless decomposition}\label{sect:TTFull}

In the first part of this section, we have derived the transverse traceless decomposition for the tracefree hook antisymmetric part of the torsion. The decomposition for the purely vectorial part is given by the Hodge (or Helmoltz) decomposition. Finally, the result for the completely antisymmetric part remains identical to the one given in the previous section Eq.\ \eref{eq:H-decomp}. Therefore, we can now summarize these results in the final form of the transverse traceless decomposition of the torsion, which is one of our main results
\begin{eqnarray}\label{eq:TTdecomp-full}
T^\rho{}_{\mu\nu} & = \frac{1}{d-1} \left( \delta^\rho{}_\nu \tau_\mu + \delta^\rho{}_\nu \partial_\mu \varphi - \delta^\rho{}_\mu \tau_\nu - \delta^\rho{}_\mu \partial_\nu \varphi \right)\\
&\quad + \frac{1}{3!(d-3)!} \varepsilon^{\sigma_1\dots\sigma_{d-3}\rho}{}_{\mu\nu} \left( \theta_{\sigma_1\dots\sigma_{d-3}} + \nabla_{[\sigma_1} \pi_{\sigma_2\dots\sigma_{d-3}]} \right)\nonumber \\
& \quad + \kappa^\rho{}_{\mu\nu} + \nabla_\mu S^\rho{}_\nu - \nabla_\nu S^\rho{}_\mu + 2 \nabla^\rho A_{\mu\nu} + \nabla_\mu A^\rho{}_\nu - \nabla_\nu A^\rho{}_\mu + \nabla^\rho \nabla_\mu \zeta_\nu - \nabla^\rho \nabla_\nu \zeta_\mu\nonumber\\
& \quad  - \frac{1}{d-1} \left[ \delta^\rho{}_\nu \left( R^\lambda{}_\mu \zeta_\lambda - \square \zeta_\mu \right) - \delta^\rho{}_\mu \left( R^\lambda{}_\nu \zeta_\lambda - \square \zeta_\nu \right) \right] \, ,\nonumber
\end{eqnarray}
where all the tensors on the r.h.s.\ are transverse and traceless. The counting of the degrees of freedom carried by each tensor is shown in Table \ref{tab:theonlytable}. In particular, we have exploited the flat-space limit to get rid of the spurious vector mode.

In the first technical part of this section, we have proved that this transverse traceless decomposition exists and is unique on closed manifolds. Indeed, in this case, we know that the possible existence of zero-modes of the traceless Curtright forms does not prevent the inversion of the second-order coupled differential equations whose solutions are the longitudinal modes. Moreover, we have also shown that on maximally symmetric spaces these equations decouple, and the task of finding the explicit solutions for the longitudinal modes simplifies considerably. We stress that, in deriving our decomposition, we have assumed the background affine-connection to be Riemannian. Therefore, we cannot use Eq.\ \eref{eq:TTdecomp-full} to integrate out the torsion about some non-trivial post-Riemannian background.

As can be seen from Table \ref{tab:theonlytable}, some of the modes vanish in lower dimensions. In $d=3$ we have that ${\kappa}_\perp$, ${A}_\perp$,  $\pi_{\perp}$ disappear from the spectrum. Therefore, $\kappa$ propagates a total of $5$ degrees of freedom, while $\tau_\mu$ adds up to $3$ degrees of freedom and $\theta$ becomes a pseudoscalar. In $d=2$, the only surviving modes are the two degrees of freedom of $\tau$. This can be better noticed from the fact that the degrees of freedom of $\kappa$ sum up to $\frac{1}{3}d(d+2)(d-2)$. However, from our decomposition it looks like ${\kappa}_{\perp}$ propagates one negative degree of freedom compensated by one (positive) mode of $\zeta_{\perp}$. Even though such an effect is clearly an artifact of an analytic continuation pushed beyond its validity, it would be interesting to inspect whether the limit $d\to 2$ of some dynamical theory of the torsion would exhibit a spectrum with a ghost-like mode for ${\kappa}_{\perp}$ and one mode for $\zeta_{\perp}$, and what would this imply for the corresponding quantum theory.\footnote{Such an approach would resemble what is sometimes done in two-dimensional quantum gravity, where a similar discontinuity of the degrees of freedom happens for the metric \cite{Jack:1990ey, Aida:1996zn, Martini:2021lcx, Martini:2021slj}.}

\begin{center}
	\renewcommand{\arraystretch}{2}\hspace*{-43px}
	\begin{tabular}{|c | c || c | c | c | c || c | c || c | c |}
		\cline{2-10}

		\multicolumn{1}{c|}{$\quad$} 	& \rm{$GL(d)$ irreps} &\multicolumn{4}{c ||}{$\kappa^\rho{}_{\mu\nu}$} & \multicolumn{2}{c ||}{$\tau_\mu$} & \multicolumn{2}{c|}{$\theta_{\mu_1\dots\mu_{d-3}}$}\\
		
		\cline{2-10}
		\hline
		& Modes& ${{\kappa}_{\perp}}^\rho{}_{\mu\nu}$ & $\overline{S}_{\perp}{}^{\mu\nu}$ & ${A}_{\perp}{}^{\mu\nu}$ & $\zeta_\perp{}^\mu$ & $\tau_\perp{}^\mu$ & $\varphi$ & $\theta_{\perp}{}_{\mu_1\dots\mu_{d-3}}$ & $\pi_{\perp}{}_{\mu_1\dots\mu_{d-4}}$\\
		\cline{2-10}
		$d=4$ & $J^p$ & $2^-$ & $2^+$ & $1^+$ & $1^-$ & $1^-$ & $0^+$ & $1^+$ & $0^-$\\
		\cline{2-10}
		& d.o.f. & $5$ & $5$ & $3$ & $3$ & $3$ & $1$ & $3$ & $1$\\
		
		\hline
		$d\ge 4$  & d.o.f. & $\frac{(d+1)(d-1)(d-3)}{3}$ & $\frac{(d-2)(d+1)}{2}$ & $\frac{d(d-3)}{2}+1$ & $d-1$ & $d-1$ & $1$ & $\frac{(d-1)(d-2)}{2}$ & $\frac{(d-1)(d-2)(d-3)}{6}$\\
		
		\hline
		
	\end{tabular}
	\captionof{table}{The table presents a recap of the transverse traceless decomposition of the torsion, where we have reinserted the appropriate notation to highlight the tracelessness and transversality of the tensors. For the physical case of $d=4$ we indicate both the number of degrees of freedom that are propagated and the corresponding spin-parity representation carried by the asymptotic state. In the last row we show the number of degrees of freedom of each mode in generic $d\geq 4$. }
	\label{tab:theonlytable}
\end{center}

\subsection{Relation between the two decompositions}

Here we briefly comment on the relations between the two decompositions of the hook antisymmetric part of the torsion. First, we prove that $\overline{t}_{\tiny \perp}{}^\rho{}_{\mu\nu}=(\overline{t}_{\tiny \perp}{}^\rho{}_{\mu\nu})_{\tiny \perp}$. Thus, suppose we have
\begin{eqnarray}
 \overline{t}_{\tiny \perp}{}^\rho{}_{\mu\nu} = \left( \overline{t}_{\tiny \perp}{}^\rho{}_{\mu\nu} \right)_{\tiny \perp} + (L_1 S_1)^\rho{}_{\mu\nu} + (L_2 A_1)^\rho{}_{\mu\nu} \, ,
\end{eqnarray}
where $A_1$ and $S_1$ are some antisymmetric and symmetric $2$-tensors. Taking the two independent divergences of the previous equation we find
\begin{equation}
\label{eq:system1}
\left\{
\begin{matrix}
 \nabla_\rho (L_1 S_1)^\rho{}_{\mu\nu} + \nabla_\rho (L_2 A_1)^\rho{}_{\mu\nu} = 0 \,,\\
 \nabla_\mu (L_1 S_1)^{(\rho|\mu|\nu)} + \nabla_\mu (L_2 A_1)^{(\rho|\mu|\nu)} = 0\,.
 \end{matrix}
 \right.
\end{equation}
As before, one can consider the quadratic form given by the longitudinal operators and integrate by parts to show that
\begin{eqnarray}
&\int \sqrt{g}\,\left[(L_1 S)+(L_2 A)\right]^\rho{}_{\mu\nu}\left[(L_1 S)+(L_2 A)\right]_\rho{}^{\mu\nu}=\\
&\qquad\qquad-\int \sqrt{g}\left[S_{\rho\nu}\,\left(div_2\left(L_1S+L_2A\right)\right)^{\rho\nu}+3A_{\mu\nu}\,\left(div_1\left(L_1S+L_2A\right)\right)^{\mu\nu}\right]\,.\nonumber
\end{eqnarray}
Therefore, the kernel of the system of differential equations \eref{eq:system1} coincides with that of $(L_1 S_1)^\rho{}_{\mu\nu} + (L_2 A_1)^\rho{}_{\mu\nu}$.

Now we show that $\overline{(t_{\tiny \perp}{}^\rho{}_{\mu\nu})}{}_{\tiny \perp} = \overline{t}_{\tiny \perp}{}^\rho{}_{\mu\nu}$. In the following we shall denote with $A_i$ and $S_i$, $i=1,2,3$, some antisymmetric and symmetric $2$-tensors, respectively. First, for the transverse part of $t$ we can write
\begin{eqnarray}\label{eq:rel1}
 (t_{\tiny \perp})^\rho{}_{\mu\nu} &= \overline{((t_{\tiny \perp})^\rho{}_{\mu\nu})}{}_{\tiny \perp} + (M_1 S_1)^\rho{}_{\mu\nu} + (M_2 A_1)^\rho{}_{\mu\nu}\\
  &\quad+ \frac{1}{d-1} \left( \delta^\rho{}_\nu (t_{\tiny \perp})^\lambda{}_{\mu\lambda} - \delta^\rho{}_\mu (t_{\tiny \perp})^\lambda{}_{\nu\lambda} \right) \, ,\nonumber
\end{eqnarray}
Moreover, we can also write down the transverse decomposition of $t$, i.e.,
\begin{eqnarray}\label{eq:rel2}
 t^\rho{}_{\mu\nu} = t_{\tiny \perp}{}^\rho{}_{\mu\nu} + (L_1 S_2)^\rho{}_{\mu\nu} + (L_2 A_2)^\rho{}_{\mu\nu} \, .
\end{eqnarray}
Notice that the two following equations, which concern the trace of $t_{\tiny \perp}$ and the relation between the Curtright and tracefree Curtright forms, hold
 \begin{eqnarray}
   &(t_{\tiny \perp})^\lambda{}_{\mu\lambda} = t^\lambda{}_{\mu\lambda} - (L_1 S_2)^\lambda{}_{\mu\lambda} - (L_2 A_2)^\lambda{}_{\mu\lambda} \, , \\
   &(M_1 S_2)^\rho{}_{\mu\nu} + (M_2 A_2)^\rho{}_{\mu\nu} = (L_1 S_2)^\rho{}_{\mu\nu} + (L_2 A_2)^\rho{}_{\mu\nu}\\
  &\qquad\qquad - \frac{1}{d-1} \left[ \delta^\rho{}_\nu \left(  t^\lambda{}_{\mu\lambda} - (t_{\tiny \perp})^\lambda{}_{\mu\lambda} \right) - \delta^\rho{}_\mu \left(  t^\lambda{}_{\nu\lambda} - (t_{\tiny \perp})^\lambda{}_{\nu\lambda} \right) \right] \, .\nonumber
 \end{eqnarray}
By plugging Eq.\ \eref{eq:rel1} into Eq.\ \eref{eq:rel2} and exploiting the previous two equations we find
\begin{eqnarray}
 t^\rho{}_{\mu\nu} =& \overline{((t_{\tiny \perp})^\rho{}_{\mu\nu})}{}_{\tiny \perp} + (M_1 (S_1 + S_2))^\rho{}_{\mu\nu} + (M_2 (A_1 + A_2))^\rho{}_{\mu\nu}\\
 & + \frac{1}{d-1} \left( \delta^\rho{}_\nu t^\lambda{}_{\mu\lambda} - \delta^\rho{}_\mu t^\lambda{}_{\nu\lambda} \right) \, .\nonumber
\end{eqnarray}
On the other hand, the transverse traceless decomposition reads
\begin{eqnarray}
 t^\rho{}_{\mu\nu} = \overline{t}_{\tiny \perp}{}^\rho{}_{\mu\nu} + (M_1 S_3)^\rho{}_{\mu\nu} + (M_2 A_3)^\rho{}_{\mu\nu} 
 + \frac{1}{d-1} \left( \delta^\rho{}_\nu t^\lambda{}_{\mu\lambda} - \delta^\rho{}_\mu t^\lambda{}_{\nu\lambda} \right) \, .
\end{eqnarray}
By equating the last two equations we find
\begin{eqnarray}
 \overline{t}_{\tiny \perp}{}^\rho{}_{\mu\nu} - \overline{((t_{\tiny \perp})^\rho{}_{\mu\nu})}{}_{\tiny \perp} = (M_1 (S_1 + S_2 - S_3))^\rho{}_{\mu\nu} + (M_2 (A_1 + A_2 - A_3))^\rho{}_{\mu\nu} \, .
\end{eqnarray}
The previous equation is an equality between a transverse traceless and a longitudinal tensor. Thus, using the orthogonality between such tensors, we see that the only way this relation can hold is when both sides vanish identically. This proves that $\overline{t}_{\tiny \perp}{}^\rho{}_{\mu\nu}=\overline{((t_{\tiny \perp})^\rho{}_{\mu\nu})}{}_{\tiny \perp}$. Thus, we have a commutative diagram $(\perp) \circ (\overline{\perp}) = (\overline{\perp}) \circ (\perp)$ on hook antisymmetric tensors.

\section{Wheeler-DeWitt supermetric and the functional measure}\label{sect:WD&FM}

It is well known that the functional integral over a given tensor field can be formally defined through the normalization of the Gau{\ss}ian integral \cite{Mazur:1990ak,Mottola:1995sj,Percacci:2017fkn}
\begin{eqnarray}\label{eq:def-path-int-phi}
 \int D \phi \exp \left\{ - \frac{1}{2} \phi^{T} G \phi \right\} = 1 \, ,
\end{eqnarray}
where $\phi$ is a generic tensor or spinor field, while $G$ is the Wheeler-DeWitt supermetric, which is required to be ultra-local and hermitian. The previous equation defines the functional measure of the tensor field $\phi$. In the presence of some gauge invariance, this definition is equivalent to the usual Faddeev-Popov trick, where we manifestly integrate over the volume of the gauge group \cite{Mottola:1995sj}. Moreover, such a formal definition is usually accompanied by a covariant decomposition of the given tensor field $\phi$ into its spin-parity eigenstates. Nevertheless, regardless of the possibility of having a gauge structure, Eq.\ \eref{eq:def-path-int-phi} implicitly gives the form of the Jacobian factor which arises when we make the change of variables from $\phi$ to its spin-parity eigenmodes
\begin{eqnarray}
 D \phi = J (D \overline{\phi}_{\tiny \perp}) \dots \, ,
\end{eqnarray}
where the dots extend over all the longitudinal and trace modes, while $\overline{\phi}_{\tiny \perp}$ is both transverse and traceless.

In the following, we shall first study a convenient form of the Wheeler-DeWitt supermetric for the torsion, making contact with the MAG literature for $d=4$. Then we will substitute the transverse-traceless decomposition found in the previous section Eq.\ \eref{eq:TTdecomp-full} into the definition of the functional measure. We will be able to give a particularly simple form of the Jacobian factor in the limit of maximally symmetric background Riemannian geometries.

\subsection{Wheeler-DeWitt supermetric}

As we have explained above, once a proper Wheeler-DeWitt supermetric is given, we can define the functional measure for torsion perturbations by
\begin{eqnarray}\label{def-int-measure}
 \int D T^\rho{}_{\mu\nu} \exp \left\{ -  \int \sqrt{g} \, T^\lambda{}_{\alpha\beta} \, G_\lambda{}^{\alpha\beta}{}_\rho{}^{\mu\nu} \, T^\rho{}_{\mu\nu} \right\} = 1 \, .
\end{eqnarray}
Requiring, as usual, such super-metric to be ultra-local, to share the same symmetries of the torsion tensor over each of the two sets of indices, and to be invariant under the exchange of them, we find
\begin{eqnarray}
\label{eq:supermetric}
 G_\lambda{}^{\alpha\beta}{}_\rho{}^{\mu\nu} = & \frac{A}{2} \left[ g_{\lambda\rho} \left( g^{\alpha\mu} g^{\beta\nu} - g^{\alpha\nu} g^{\beta\mu} \right)\right.\\
 & + B \left( \delta^\nu{}_\rho g^{\alpha\mu} \delta^\beta{}_\lambda - \delta^\mu{}_\rho g^{\alpha\nu} \delta^\beta{}_\lambda - \delta^\nu{}_\rho g^{\beta\mu} \delta^\alpha{}_\lambda + \delta^\mu{}_\rho g^{\beta\nu} \delta^\alpha{}_\lambda \right) \nonumber\\
 & + \left. C \varepsilon_{\sigma_1\dots\sigma_{d-3}\lambda}{}^{\alpha\beta} \varepsilon^{\sigma_1\dots\sigma_{d-3}}{}_\rho{}^{\mu\nu} \right] \, ,\nonumber
\end{eqnarray}
for $A$, $B$, and $C$ some real coefficients.
Let us note that, since the Weyl weight of the torsion tensor is bound to be zero, as it is for all the tensor components of any affine-connection, in $d\neq2$ the previous definition of the functional measure cannot be Weyl invariant.

Of course, there exist other possible parametrizations of the supermetric.
For example, one can consider all the independent scalar contractions of two copies of the torsion tensor (see Eq.\eref{WD-telep} below) and take the second variation with respect to $T^\rho{}_{\mu\nu}$. This defines the supermetric in the teleparallel basis, whose relation to Eq.\eref{eq:supermetric} shall be studied below.

Similarly to the case of symmetric $2$-tensors, there is an overall normalization constant that is left free. In our choice of normalization, such a constant coincides with the eigenvalue of the top-spin mode in the torsion tensor, meaning that fixing $A=1$ implies that we are normalizing all the degrees of freedom in units of $\kappa^\rho{}_{\mu\nu}$.\footnote{In other words, we discard the limit in which the supermetric reduces to a projector onto the orthogonal complement to $\kappa^\rho{}_{\mu\nu}$, since this corresponds to $A\to 0$, $B\to\infty$, $C\to\infty$. However, other choices of basis would make this limit explicitly finite.} In order to get an insight about the role of $B$ and $C$
we decompose once more the torsion tensor into its irreducible components as
\begin{eqnarray}
 T^\rho{}_{\mu\nu} = \kappa^\rho{}_{\mu\nu} + \frac{1}{d-1} \left( \delta^\rho{}_\nu \tau_\mu - \delta^\rho{}_\mu \tau_\nu \right) + \frac{1}{3!(d-3)!} \varepsilon^{\sigma_1\dots\sigma_{d-3}\rho}{}_{\mu\nu} \theta_{\sigma_1\dots\sigma_{d-3}} \, ,
\end{eqnarray}
where the dependence of the axial-torsion part on the spacetime dimension has a well-defined meaning only for integer values of $d$. From the analogy with the Wheeler-DeWitt supermetric for symmetric $2$-tensors \cite{Mazur:1990ak,Mottola:1995sj}, we know that there are some values of $B$ and $C$ for which the supermetric becomes singular. These are found by computing the eigenvalues of the trace and completely antisymmetric components of the torsion
 \begin{eqnarray}
&\hspace*{-60px}  G_\lambda{}^{\alpha\beta}{}_\rho{}^{\mu\nu} \frac{1}{d-1} \left( \delta^\rho{}_\nu \tau_\mu - \delta^\rho{}_\mu \tau_\nu \right) =  \, A \left( 1 + B \, (d-1) \right) \frac{1}{d-1} \left( \delta^\beta{}_\lambda \tau^\alpha - \delta^\alpha{}_\lambda \tau^\beta \right) \, ;\\
 &\hspace*{-60px} G_\lambda{}^{\alpha\beta}{}_\rho{}^{\mu\nu} \frac{1}{3!(d-3)!} \varepsilon^{\sigma_1\dots\sigma_{d-3}\rho}{}_{\mu\nu} \theta_{\sigma_1\dots\sigma_{d-3}} =  \, A \left( 1 + (d-3)!\frac{C}{2} \right) \frac{1}{3!(d-3)!} \varepsilon_{\sigma_1\dots\sigma_{d-3}\lambda}{}^{\alpha\beta} \theta^{\sigma_1\dots\sigma_{d-3}} \, .
 \end{eqnarray}
Thus, the supermetric is singular for $B= - \frac{1}{d-1}$ and $C=-\frac{2}{3!(d-3)!}$, while it is positive definite for $B > - \frac{1}{d-1}$ and $C >-\frac{2}{3!(d-3)!}$. Whenever $B$ is equal to its singular value, the supermetric becomes a projector onto the trace-free vector-valued $2$-forms. The same holds \emph{mutatis mutandis} for the singular value of $C$, for which the supermetric becomes a projector onto hook antisymmetric tensors.

Let us make a final remark on the sign of these eigenvalues. It is well known that, for generic values of the couplings, in torsion Lagrangians there are tachyonic modes \cite{Sezgin:1979zf,Sezgin:1981xs,Baldazzi:2021kaf}.
One may then hope to avoid the presence of some of them through a mechanism similar to that used in Ref.\ \cite{Mottola:1995sj}, i.e., by declaring some eigenvalues to be (semi-)negative definite. This procedure would in any case be of a very complicated nature, given the huge number of spin-parity eigenstates.

\subsubsection{The $d=4$ limit and MAGs}

As we have said above, another basis for the Wheeler-DeWitt supermetric may be derived by taking into account all the non-trivial and independent contractions of two torsion tensors, and taking the (functional) derivatives of w.r.t. the torsion tensor itself twice. Here we want to provide a mapping between these two bases for the physical case of a four-dimensional spacetime. A well-known basis for such contractions lies at the origin of the teleparallel theories of Gravity \cite{BeltranJimenez:2019esp}
\begin{eqnarray}\label{WD-telep}
 \mathbb{T} = a \, T^\rho{}_{\mu\nu} T_\rho{}^{\mu\nu} + b \, T^\mu{}_{\nu\rho} T^{\rho\nu}{}_\mu + c \, T^{\lambda}{}_{\mu\lambda} T_{\nu}{}^{\mu\nu} \, .
\end{eqnarray}
Note that we obtain the teleparallel action equivalent to GR for $a=\frac{1}{4}$, $b=\frac{1}{2}$ and $c=-1$ \cite{BeltranJimenez:2019esp}. Now let us manipulate the previous expression in order to find the underlying supermetric
\begin{eqnarray}
& \hspace*{-50px}\mathbb{T} =  \, T^\lambda{}_{\alpha\beta} \left[ \, a \, g_{\lambda\rho} g^{\alpha\mu} g^{\beta\nu} + b \, \delta^\nu{}_\lambda g^{\alpha\mu} \delta^\beta{}_\rho + c \, \delta^\beta{}_\lambda \delta^\nu{}_\rho g^{\alpha\mu} \right] T^\rho{}_{\mu\nu} \, \\\nonumber
&\hspace*{-50px}\,\,\,\, =  \, T^\lambda{}_{\alpha\beta} \left[ \frac{a}{2} \, g_{\lambda\rho} \left( g^{\alpha\mu} g^{\beta\nu} - g^{\alpha\nu} g^{\beta\mu} \right) + \frac{b}{4} \left( \delta^\nu{}_\lambda g^{\alpha\mu} \delta^\beta{}_\rho - \delta^\mu{}_\lambda g^{\alpha\nu} \delta^\beta{}_\rho - \delta^\nu{}_\lambda g^{\beta\mu} \delta^\alpha{}_\rho + \delta^\mu{}_\lambda g^{\beta\nu} \delta^\alpha{}_\rho \right) \right.\\ \nonumber
 &\hspace*{-50px} \qquad \qquad + \left. \frac{c}{4} \left( \delta^\beta{}_\lambda \delta^\nu{}_\rho g^{\alpha\mu} - \delta^\beta{}_\lambda \delta^\mu{}_\rho g^{\alpha\nu} - \delta^\alpha{}_\lambda \delta^\nu{}_\rho g^{\beta\mu} + \delta^\alpha{}_\lambda \delta^\mu{}_\rho g^{\beta\nu} \right) \right] T^\rho{}_{\mu\nu} \, .
\end{eqnarray}
Exploiting the fact that	
\begin{eqnarray}
 \hspace*{-50px}\varepsilon_{\sigma\lambda}{}^{\alpha\beta} \varepsilon^\sigma{}_\rho{}^{\mu\nu} = g_{\lambda\rho} g^{\alpha\mu} g^{\beta\nu} - g_{\lambda\rho} g^{\alpha\nu} g^{\beta\mu} + \delta^\alpha{}_\rho g^{\beta\mu} \delta^\nu{}_\lambda - \delta^\alpha{}_\rho g^{\beta\nu} \delta^\mu{}_\lambda + \delta^\beta{}_\rho \delta^\mu{}_\lambda g^{\nu\alpha} - \delta^\beta{}_\rho \delta^\nu{}_\lambda g^{\mu\alpha} \, ,
\end{eqnarray}
we can read off the mappings between the two bases for the Wheeler-DeWitt supermetric
\begin{eqnarray}
a = A ( 1 + C) \,,\qquad
b = -2 AC \,,\qquad
c = 2 A B \,.
\end{eqnarray}
In terms of these coefficients the definiteness of the supermetric reads
\begin{eqnarray}
 a> \frac{2}{3} \, , \qquad b < \frac{2}{3} \, , \qquad c> - \frac{2}{3} \, ,
\end{eqnarray}
where we have chosen to fix the overall normalization constant $A$ to be one.

\subsection{The functional measure and the Jacobian}

Having established a convenient form of the supermetric, we turn to the formal definition of the functional measure. As we have explained above, this is usually done by requiring the normalization of the Gau{\ss}ian integral, see Eq.~\eref{eq:def-path-int-phi}. At this point we exploit our covariant decomposition of the torsion into its spin-parity eigenstates Eq.\ \eref{eq:TTdecomp-full} and split the functional measure as
\begin{eqnarray}\label{eq::decomp-jacobian}
 D T^\rho{}_{\mu\nu} = J  \, D \tau_\mu \, D \varphi \, D \theta_{\mu_1\dots\mu_{d-3}} \, D \pi_{\mu_1\dots\mu_{d-4}} \, D \kappa^\rho{}_{\mu\nu} \, D S^\rho{}_\nu \, D A_{\mu\nu} \, D \zeta_\mu \, ,
\end{eqnarray}
where $J$ is the Jacobian factor which arises from the change of variables and all the tensors which will appear from now on are assumed to be transverse.

Let us expand the exponential in Eq.\ \eref{def-int-measure}: we start by taking take care of the action of the Wheeler-DeWitt supermetric $G$ on the full torsion tensor $T$
\begin{eqnarray}
\hspace*{-50px}(GT)^\rho{}_{\mu\nu} = & \,  A \left[\left( B + \frac{1}{d-1} \right) \left( \delta^\rho{}_\nu \tau_\mu + \delta^\rho{}_\nu \partial_\mu \varphi - \delta^\rho{}_\mu \tau_\nu - \delta^\rho{}_\mu \partial_\nu \varphi \right)\right. \\
& \, + \frac{1}{6}\left(  \frac{C}{2} + \frac{1}{(d-3)!} \right) \varepsilon^{\sigma_1\dots\sigma_{d-3}\rho}{}_{\mu\nu} \left( \theta_{\sigma_1\dots\sigma_{d-3}} + \nabla_{[\sigma_1} \pi_{\sigma_2\dots\sigma_{d-3}]} \right) \nonumber\\
 & \, \left. +  \, \kappa^\rho{}_{\mu\nu} + \nabla_\mu S^\rho{}_\nu - \nabla_\nu S^\rho{}_\mu + 2 \nabla^\rho{} A_{\mu\nu} + \nabla_\mu A^\rho{}_\nu - \nabla_\nu A^\rho{}_\mu + \nabla^\rho \nabla_\mu \zeta_\nu - \nabla^\rho \nabla_\nu \zeta_\mu \right. \nonumber\\
 & \, \left. -  \frac{1}{d-1} \left( \delta^\rho{}_\nu \left( R^\lambda{}_\mu \zeta_\lambda - \square \zeta_\mu \right) - \delta^\rho{}_\mu \left( R^\lambda{}_\mu \zeta_\lambda - \square \zeta_\mu  \right) \right) \right] \, .\nonumber
\end{eqnarray}
Now we contract the former expression with the decomposition of $T^\rho{}_{\mu\nu}$ given in Eq.~\eref{eq:TTdecomp-full} and integrate the resulting scalar. Since the background connection is Riemannian we are allowed to integrate by parts. Therefore, exploiting integration by parts and the transversality properties of the tensors, we obtain
\begin{eqnarray}
&\hspace*{-60px}\int \sqrt{g} T G T =  \, \int \sqrt{g} A \Bigg\{ 2\left( B + \frac{1}{d-1} \right) ( \tau_\mu \tau^\mu - \varphi \square \varphi )  + \kappa^\rho{}_{\mu\nu} \kappa_\rho{}^{\mu\nu} \\
&\hspace*{-50px}\, + \frac{1}{3!(d-3)!}\left( \frac{C}{2} + \frac{1}{(d-3)!} \right) \Bigg[\theta_{\mu_1\dots\mu_{d-3}} \theta^{\mu_1\dots\mu_{d-3}}- \frac{1}{d-3}\pi_{\mu_1\dots\mu_{d-4}} \square \pi^{\mu_1\dots\mu_{d-4}} \nonumber\\
&\hspace*{-50px} +\frac{d-4}{d-3}\, \Bigg(\pi_{\mu\mu_{1}\dots\mu_{d-5}}R^\mu{}_\nu \pi^{\nu\mu_{1}\dots\mu_{d-5}}- \frac{d-5}{2} \pi_{\mu\nu\mu_{2}\dots\mu_{d-5}}R^{\mu\nu}{}_{\rho\sigma} \pi^{\rho\sigma\mu_{2}\dots\mu_{d-5}} \Bigg) \Bigg]\nonumber\\
&\hspace*{-50px} \, -2 S^{\mu\nu} \square S_{\mu\nu} - 2 S^{\alpha\beta} R_{\alpha\mu\beta\nu} S^{\mu\nu} + 2 S_\rho{}^\mu R_{\mu\nu} S^{\rho\nu} - 6 A^{\mu\nu} \square A_{\mu\nu} - 6 A_\rho{}^\nu R_{\lambda\nu} A^{\rho\lambda} + 3 A^{\alpha\beta} R_{\alpha\beta\mu\nu} A^{\mu\nu}\nonumber \\
&\hspace*{-50px} \, + 2 \frac{d-2}{d-1} \zeta^\mu \square^2 \zeta_\mu - 2 \frac{d-3}{d-1} \zeta_\mu R^{\mu\nu} \square \zeta_\nu - \frac{2d}{d-1} \zeta_\mu R^\mu{}_\lambda R^{\lambda\nu} \zeta_\nu - \frac{1}{2} \zeta^2 \square R + 2 \zeta_\nu R_{\lambda\mu} \nabla^\mu \nabla^\lambda \zeta^\nu \nonumber\\
&\hspace*{-50px} \, - 2 \zeta_\nu (\nabla_\lambda R^\nu{}_\mu) \nabla^\lambda \zeta^\mu - 4 \zeta_\nu R^{\rho\nu\mu\lambda} \nabla_\rho \nabla_\mu \zeta_\lambda - 12 A^\mu{}_\nu R_{\lambda\mu} S^{\lambda\nu} + 4 R_\lambda{}^\mu S^{\lambda\nu} ( \nabla_\nu \zeta_\mu  - \nabla_\mu \zeta_\nu ) \nonumber\\
&\hspace*{-50px} \, + 12 (\nabla_\mu \zeta_\nu ) R^{\mu\nu\rho\sigma} A_{\rho\sigma} - 3 A_{\mu\nu} R^{\mu\lambda} \nabla_\lambda \zeta^\nu + 20 A_{\mu\lambda} R^{\lambda\nu} \nabla^\mu \zeta_\nu \Bigg\} \, .\nonumber
\end{eqnarray}
We see that there is a striking difference from the usual York decomposition. Now two kinetic terms explicitly depend on the vector and pseudo-vector traces factors $B$ and $C$. Moreover, the last line contains all the mixing terms among the three longitudinal modes of the purely tensorial torsion. We notice that, while the $A$-$S$ and $\zeta$-$S$ couplings disappear in the Einstein-manifold limit, this is not the case for the $\zeta$-$A$ coupling.

As it is usually done when dealing with the functional measure over metrics \cite{Mazur:1990ak,Mottola:1995sj}, we specialize to an Einstein manifold, i.e., $R_{\mu\nu}=\frac{1}{d} g_{\mu\nu} R$. In such spacetimes, the Weyl tensor can be written as
\begin{eqnarray}
 C_{\mu\nu\rho\sigma} = R_{\mu\nu\rho\sigma} - \frac{R}{d(d-1)} \left( g_{\mu\rho} g_{\nu\sigma} - g_{\mu\sigma} g_{\nu\rho} \right) \, .
\end{eqnarray}
Exploiting such relation, as well as the transversality of the tensors and the contracted form of the second Bianchi identity (we are tacitly assuming $d \neq 2$), we get
\begin{eqnarray}
&\hspace*{-60px}\int \sqrt{g} T G T = \, \int \sqrt{g} A \Bigg\{2\left( B + \frac{1}{d-1} \right) ( \tau_\mu \tau^\mu - \varphi \square \varphi )  + \kappa^\rho{}_{\mu\nu} \kappa_\rho{}^{\mu\nu} \\
&\hspace*{-50px}\, \, + \frac{1}{3!(d-3)!}\left( \frac{C}{2} + \frac{1}{(d-3)!} \right) \Bigg[\theta_{\mu_1\dots\mu_{d-3}} \theta^{\mu_1\dots\mu_{d-3}}- \frac{1}{d-3}\pi_{\mu_1\dots\mu_{d-4}} \square \pi^{\mu_1\dots\mu_{d-4}} \nonumber\\
&\hspace*{-50px} \, +\frac{d-4}{d-3}\, \Bigg( \frac{4}{d(d-1)} \pi_{\mu_{1}\dots\mu_{d-4}} R \pi^{\mu_{1}\dots\mu_{d-4}}- \frac{d-5}{2} \pi_{\mu\nu\mu_{2}\dots\mu_{d-5}}C^{\mu\nu}{}_{\rho\sigma} \pi^{\rho\sigma\mu_{2}\dots\mu_{d-5}} \Bigg) \Bigg]\nonumber\\
&\hspace*{-50px} \, -2 S^{\mu\nu} \square S_{\mu\nu} - 2 S^{\alpha\beta} C_{\alpha\mu\beta\nu} S^{\mu\nu} + \frac{2}{d-1} R S^{\mu\nu} S_{\mu\nu} - 6 A^{\mu\nu} \square A_{\mu\nu} - \frac{6(d-2)}{d(d-1)} R A^{\mu\nu} A_{\mu\nu} + 3 A^{\alpha\beta} C_{\alpha\beta\mu\nu} A^{\mu\nu} \nonumber\\
&\hspace*{-50px} \, + 2 \frac{d-2}{d-1} \zeta^\mu \square^2 \zeta_\mu -\frac{d-2}{2d} \zeta^2 \square R - \frac{2(d-2)}{d^2 (d-1)} R^2 \zeta^\mu \zeta_\mu + \frac{8(d+2)}{d(d-1)} A_{\mu\nu} R \nabla^\mu \zeta^\nu\nonumber \\
&\hspace*{-50px} \, + 12 A^{\mu\nu} C_{\mu\nu\rho\sigma} \nabla^\rho \zeta^\sigma + 4 C_{\mu\nu\rho\sigma} \zeta^\mu \nabla^\nu \nabla^\rho \zeta^\sigma
  \Bigg\} \, . \nonumber
\end{eqnarray}
Let us observe that, despite our simplifying assumption of dealing with Einstein manifolds, we have not been able to diagonalize the Hessian in field space. This feature clearly distinguishes the functional integration over torsion fluctuations from what happens when integrating over metric perturbations \cite{Mazur:1990ak,Mottola:1995sj,Percacci:2017fkn}. Thus, to gain some further insight, we specialize to maximally symmetric spaces, i.e.,  $C_{\mu\nu\rho\sigma}=0$ and $\nabla_\mu R=0$. In such a limit we obtain
\begin{eqnarray}
&\hspace*{-60px} \int \sqrt{g} T G T =  \, \int \sqrt{g} A \left\{2\left( B + \frac{1}{d-1} \right) ( \tau_\mu \tau^\mu - \varphi \square \varphi )  + \kappa^\rho{}_{\mu\nu} \kappa_\rho{}^{\mu\nu} \right. \\
&\hspace*{-50px}\, + \frac{1}{3!(d-3)!}\left( \frac{C}{2} + \frac{1}{(d-3)!} \right) \Bigg[\theta_{\mu_1\dots\mu_{d-3}} \theta^{\mu_1\dots\mu_{d-3}}- \frac{1}{d-3}\pi_{\mu_1\dots\mu_{d-4}} \square \pi^{\mu_1\dots\mu_{d-4}}\nonumber\\
	&\hspace*{-50px}+ \frac{4(d-4)}{d(d-1)(d-3)} \pi_{\mu_{1}\dots\mu_{d-4}} R \pi^{\mu_{1}\dots\mu_{d-4}} \Bigg]\nonumber\\
 &\hspace*{-50px} \, -2 S^{\mu\nu} \square S_{\mu\nu} + \frac{2}{d-1} S_{\mu\nu} R \, S^{\mu\nu} - 6 A^{\mu\nu} \square A_{\mu\nu} - \frac{6(d-2)}{d(d-1)} A^{\mu\nu} R \, A_{\mu\nu} \nonumber\\
 &\hspace*{-50px}\left.+ 2 \frac{d-2}{d-1} \zeta^\mu \square^2 \zeta_\mu - \frac{2(d-2)}{d^2(d-1)} \zeta_\mu R^2 \zeta^\mu  \right\} \, . \nonumber
\end{eqnarray}
We see that the requirement of maximal symmetry has diagonalized the Hessian found by taking the functional variations w.r.t.\ to the various fields. Therefore, the normalization of the functional integral now takes a remarkably simple form
\begin{eqnarray}
\label{eq:gaussIntegr}
&\hspace*{-60px} 1 = \int  D T \exp \left\{ - \frac{1}{2} TGT \right\} = \\\nonumber
 &\hspace*{-54px} =\int J \, D\tau \, D \varphi \, D \theta \, D \pi \, D \kappa \,  D S \,  D A \,  D \zeta \,\times\nonumber\\
 &\hspace*{-55px}\quad\times \exp\left\{ \,  - \frac{1}{2} \left( \tau \Delta_{\tiny \tau} \tau + \varphi \Delta_{\tiny \varphi} \varphi + \theta \Delta_{\tiny \theta} \theta + \pi \Delta_{\tiny \pi} \pi + \kappa \Delta_{\tiny \kappa} \kappa + S \Delta_{\tiny S} S + A \Delta_{\tiny A} A + \zeta \Delta_{\tiny \zeta} \zeta \right) \right\} \, ,\nonumber
\end{eqnarray}
where the explicit expressions of the non-trivial differential operators are
 \begin{eqnarray}
  &\hspace*{-50px}\varphi \Delta_{\tiny \varphi} \varphi  = - 2\left( B + \frac{1}{d-1} \right) \varphi \square \varphi \, , \label{Delta-pi}\\
  &\hspace*{-50px}\pi \Delta_{\tiny \pi} \pi  = - \frac{1}{3!(d-3)!(d-3)}\left( \frac{C}{2} + \frac{1}{(d-3)!} \right) \left[ \pi_{\mu_1\dots\mu_{d-4}} \square \pi^{\mu_1\dots\mu_{d-4}}\right.\\
 &\hspace*{-50px}\qquad\quad\left. - \frac{4(d-4)}{d(d-1)} \pi_{\mu_{1}\dots\mu_{d-4}} R \pi^{\mu_{1}\dots\mu_{d-4}} \right] \, , \nonumber\\
 &\hspace*{-50px} S \Delta_{\tiny S} S  = -2 S^{\mu\nu} \square S_{\mu\nu} + \frac{2}{d-1} S_{\mu\nu} R \, S^{\mu\nu} \, , \\
&\hspace*{-50px}  A \Delta_{\tiny A} A  = - 6 A^{\mu\nu} \square A_{\mu\nu} - \frac{6(d-2)}{d(d-1)} A^{\mu\nu} R \, A_{\mu\nu} \, , \\
&\hspace*{-50px}  \zeta \Delta_{\tiny \zeta} \zeta  =  2 \frac{d-2}{d-1} \zeta^\mu\square^2 \zeta_\mu - \frac{2(d-2)}{d^2(d-1)} \zeta_\mu R^2 \zeta^\mu \, .\label{eq:DeltaA}
 \end{eqnarray}
Note that all the above operators are of Laplace type; however, $\Delta_{\tiny \zeta}$ is a higher derivative operator of quartic order. Assuming the validity of Fubini's theorem and requiring the usual normalization Eq.\ \eref{eq:def-path-int-phi} for all the transverse fields, we can read off the formal expression of the functional Jacobian
\begin{eqnarray}\label{eq:Jacobian-max-sym}
\hspace*{-50px} J = \left[{\det}{'}_\varphi (- \square) \right]^{\frac{1}{2}} \left[ {\det}_\pi \left(- \square + \frac{4(d-4) \, R}{d(d-1)}\right) \right]^{\frac{1}{2}} \left[ {\det}_S \left(- \square + \frac{R}{d-1}\right) \right]^{\frac{1}{2}}\times\\
\hspace*{-50px}\qquad\times \left[ {\det}_A \left(- \square - \frac{(d-2) \, R}{d(d-1)} \right) \right]^{\frac{1}{2}} \left[ {\det}_\zeta \left(\square^2 - \frac{R^2}{d^2}\right) \right]^{\frac{1}{2}} \, .\nonumber
\end{eqnarray}
The prime in the previous equation stands for integration over non-constant configurations since constant ones never contribute to the Jacobian. Notice that also the axial term would need a prime in $d=4$.

This Jacobian would suggest a non-local redefinition of the longitudinal modes, in analogy with what happens when we analyze the path-integral over metric perturbations \cite{Mazur:1990ak,Mottola:1995sj,Percacci:2017fkn}. However, we know that such non-local field redefinitions are allowed only for non-physical fields which either parameterize gauge orbits or are non-dynamical ghosts \cite{Percacci:2017fkn}. Therefore, one may take into account gauge transformations of the torsion which would gauge away (some of the) longitudinal modes, see e.g.\ \cite{Curtright:1980yk}. In this case, it would be possible to perform non-local field redefinitions and the Jacobian factors would be interpreted as the (infinite) volumes of the gauge groups.

\subsection{Evaluation of the functional Jacobian on $S^4$}

In the following we shall study the previous Jacobian Eq.\ \eref{eq:Jacobian-max-sym} on a $4$-sphere.
The functional determinants which appear in Eq. \eref{eq:Jacobian-max-sym} clearly need some regularization. We choose to evaluate the divergences exploiting the heat kernel formalism, which is particularly suited to the evaluation of functional traces. We refer the reader to Chapter $14$ in \cite{Hawking:1979ig} and the more recent review \cite{Vassilevich:2003xt} for an introduction to the heat kernel formalism.
Since we are ultimately interested in the logarithmic divergences of an effective action for the torsion, defined through the measure in Eq.\ \eref{eq:gaussIntegr}, it is convenient to represent the trace of the heat kernel $K_{\Delta}(s; x, y)$ of a differential operator $\Delta$ through its Seeley-DeWitt asymptotic expansion
\begin{eqnarray}
{\rm Tr} K_{\Delta}(s) \sim \frac{1}{(4\pi s)^{\frac{d}{2}}}\sum_{n=0}^\infty B_{2n}(\Delta) s^n\,,
\end{eqnarray}
with the Seeley-DeWitt coefficients $B_{2n}(\Delta)$ being given as an integral representation in terms of the local operators $b_{2n}(\Delta)$
\begin{eqnarray}
\label{eq:bToB}
B_{2n}(\Delta) = \int{\rm d}^d x \sqrt{g}\; b_{2n}(\Delta)\,.
\end{eqnarray}
Notice that the only contribution to the logarithmic divergence of the effective action due to the functional Jacobians comes from the term with $2n=d$, which is $b_{4}(\Delta)$ for the present case.

Using the heat kernel formalism it is fairly easy to integrate out fields that are not subject to any differential constraints, i.e., tensor fields that have not been split into any covariant spin-parity decomposition. However, it is very convenient to exploit any such kind of decomposition, since one ends up having only minimal operators, as we have seen in the previous section.
%
%
%
The Seeley-DeWitt coefficients for differentially constrained rank-$k$ tensor fields can be iteratively derived from those of un-constrained rank-$j$ fields, with $j \leq k$, see \cite{Codello:2008vh,Percacci:2017fkn}. To our knowledge, the Seeley-DeWitt coefficients for transverse fields have been studied only for the physically relevant cases of $1$-forms and symmetric $2$-tensors \cite{Codello:2008vh}, while they are not known for $2$-forms.

Given a differential operator $\Delta$ acting on a tensor field, the operators acting on the longitudinal modes are found by applying $\Delta$ to the decomposition and commuting the covariant derivatives \cite{Codello:2008vh,Percacci:2017fkn}. In the simple example of the Laplacian acting on symmetric tensors, we apply $\square$ to the Killing form and find
\begin{eqnarray}
 \square \left( \nabla_\mu \xi_\nu + \nabla_\nu \xi_\mu \right) = \nabla_\mu \left( \square - \frac{d+1}{d(d-1)} R \right) \xi_\nu + \nabla_\nu \left( \square - \frac{d+1}{d(d-1)} R \right) \xi_\mu \, .
\end{eqnarray}
For closed manifolds, we have an additional subtlety to take into account, which is related to the fact that the spectra of our operators are discrete, unlike the case of a non-compact manifold. When we decompose a tensor $\phi$ into its spin-parity eigenstates, the longitudinal parts can have modes whose contribution to $\phi$ is zero. As an example consider the Killing vectors in the case of a symmetric $2$-tensor: these do not affect the symmetric $2$-tensor itself, and we need to subtract their contribution when we are computing the functional trace. On $S^4$ these Killing vectors give rise to the term proportional to $d(d+1)/2$ (their multiplicity) in the first line of the equations below. Thus, in the general case we must always subtract the contribution of these tensor modes, which will be dubbed ``spurious'' modes from now on. Fortunately, except for $A_{\mu\nu}$, such analysis has already been carried out in the case of $S^4$. Indeed, from the results of Appendix B in \cite{Codello:2008vh} specialized to the operators found in \eref{eq:Jacobian-max-sym} we readily find
\begin{eqnarray}\label{eq:func-trace1.0}
&\hspace*{-50px}{\rm Tr} \, {\rm e}^{-s\left(-\square + \frac{R}{d-1}\right)} \Big|_{\overline{S}_{\mu\nu}} =  {\rm Tr} \, {\rm e}^{-s\left(-\square + \frac{R}{d-1}\right)} \Big|_{\overline{S}_{\perp}{}_{\mu\nu}} +  {\rm Tr} \,{\rm e}^{-s\left(-\square - \frac{R}{d(d-1)}\right)} \Big|_{\xi_{\perp}{}_\mu} + {\rm Tr} \,{\rm e}^{-s\left(-\square - \frac{R}{d-1}\right)} \Big|_{\sigma}\\
&\hspace*{-50px}\qquad - {\rm e}^{s \frac{R}{d-1}} - (d+1) - \frac{d(d+1)}{2} {\rm e}^{-s\frac{(2d-1) R}{d(d-1)}} \, ; \nonumber\\
&\hspace*{-50px}{\rm Tr} \, {\rm e}^{-s\left(-\square + \frac{R}{d}\right)} \Big|_{\zeta_\mu} =  {\rm Tr} \, {\rm e}^{-s\left(-\square + \frac{R}{d}\right)} \Big|_{\zeta_{\perp}{}_{\mu}} + {\rm Tr} \,{\rm e}^{-s\left(-\square\right)} \Big|_{\phi} - 1 \, ; \\
&\hspace*{-50px}{\rm Tr} \, {\rm e}^{-s\left(-\square - \frac{R}{d}\right)} \Big|_{\zeta_\mu} =  {\rm Tr} \, {\rm e}^{-s\left(-\square - \frac{R}{d}\right)} \Big|_{\zeta_{\perp}{}_{\mu}} + {\rm Tr} \,{\rm e}^{-s\left(-\square-\frac{2R}{d}\right)} \Big|_{\phi} - {\rm e}^{\frac{2sR}{d}} \, ; \\
&\hspace*{-50px}{\rm Tr} \, {\rm e}^{-s\left(-\square\right)} \Big|_{\varphi} = {\rm Tr} \, {\rm e}^{-s\left(-\square\right)} \Big|_{\varphi^\prime} + 1 \, ; \\
&\hspace*{-50px}{\rm Tr} \, {\rm e}^{-s\left(-\square\right)} \Big|_{\pi} = {\rm Tr} \, {\rm e}^{-s\left(-\square\right)} \Big|_{\pi^\prime} + 1 \, .
\end{eqnarray}
For the antisymmetric tensor $A_{\mu\nu}$ we apply the Laplacian to its longitudinal mode, yielding
\begin{eqnarray}
 -\square \left( \nabla_\mu \xi_\nu - \nabla_\nu \xi_\mu \right) = - \nabla_\mu \left( \square + \frac{(d-3)R}{d(d-1)} \right) \xi_\nu + \nabla_\nu \left( \square + \frac{(d-3)R}{d(d-1)} \right) \xi_\mu
\end{eqnarray}
Therefore, for the trace over $A_{\mu\nu}$ we have
\begin{eqnarray}\label{eq:func-trace1.2}
{\rm Tr} \, {\rm e}^{-s\left(-\square - \frac{(d-2)R}{d(d-1)}\right)} \Big|_{A_{\mu\nu}} =  {\rm Tr} \, {\rm e}^{-s\left(-\square - \frac{(d-2)R}{d(d-1)}\right)} \Big|_{A_{\tiny \perp}{}_{\mu\nu}} + {\rm Tr} \,{\rm e}^{-s\left(-\square-\frac{(2d-5)R}{d(d-1)} \right)} \Big|_{\xi_{\tiny \perp}{}_\mu} \, .
\end{eqnarray}
Notice that, in calculating the contribution due to the longitudinal mode parametrized by $\xi_\mu$, we can consider only transverse vectors and we do not need to take into account any spurious modes. We have shown that these spurious modes are the closed $1$-forms, that, due to the triviality of the first de Rham cohomology group of the $4$-sphere \cite{Nakahara:2003nw}, are also exact forms. These can be identified with the longitudinal part of $\xi_\mu$, which can be disregarded.

Thus, from Eqs.\ \eref{eq:func-trace1.0} and \eref{eq:func-trace1.2}, we obtain the following Seeley-DeWitt coefficients
 \begin{eqnarray}
  & b_4 \left(-\square+\frac{R}{3}\right) \Big|_{\tiny \overline{S}_{\tiny \perp}{}_{\mu\nu}} = \frac{239}{432} R^2 \, ; \\
  & b_4 \left(-\square-\frac{R}{6}\right) \Big|_{\tiny A_{\tiny \perp}{}_{\mu\nu}} = \frac{109}{720} R^2 \, ; \label{eq:b4ATrans}\\
  & b_4 \left(-\square-\frac{R}{4}\right) \Big|_{\tiny \zeta_{\tiny \perp}{}_\mu} = \frac{109}{720} R^2 \, ; \\
  & b_4 \left(-\square+\frac{R}{4}\right) \Big|_{\tiny \zeta_{\tiny \perp}{}_\mu} = \frac{19}{720} R^2 \, ; \\
  & b_4 \left(-\square\right) \Big|_{\phi^\prime} = b_4 \left(-\square\right) \Big|_{\pi^\prime} = -\frac{61}{2160} R^2 \, .
 \end{eqnarray}
Summing all these contributions, we get that the behaviour of the logarithmic divergence of the functional measure for the torsion is characterized by
\begin{eqnarray}
b_4^{\rm tot} &= \frac{223}{270}R^2\,.
\end{eqnarray}
We conclude this section by noticing that instead of $S^4$, i.e., Euclidean de Sitter space, we could have considered also a $4$-dimensional hyperboloid, i.e., Euclidean Anti-de Sitter space. This choice would simplify the computations because in this case we would avoid the problem of spurious modes. However, in addition to being less physically significant from the point of view of cosmology, the Anti-de Sitter is not closed and therefore the invertibility of $\overline{\Delta}_i$, $i=2,3$ is not guaranteed.

\section{Conclusions}

In this paper we set up the machinery necessary to construct the path integral over the torsion tensor. To this end, we started by generalizing the well-known York decomposition for rank-2 symmetric tensors to hook antisymmetric tensors. Then, we applied this procedure to the full torsion tensor by writing it in terms of transverse modes only \eref{eq:TTdecomp-full}, which is one of our main results. Proceeding this way we get significant technical simplifications that, in the limit of maximally symmetric spaces, return the particularly elegant form for the Jacobian of the functional measure \eref{eq:Jacobian-max-sym}. Such decomposition does not only offer technical advantages, but it has a clear physical interpretation too. As listed in Table \ref{tab:theonlytable}, it is a decomposition in terms of the irreducible (spin) representations contained in the torsion tensor, i.e., the spin-parity eigenstates.

Let us underline that our decomposition is valid for Riemannian backgrounds only, and it would surely be important to extend our work to nontrivial torsional backgrounds. As an insight, we expect that in this more complicated case the principal symbols of the differential operators shall be the same, because the only differences arise from the endomorphism. This is crucial if one wishes to invert such operators and to uniquely obtain the rank-$2$ tensors entering our decomposition.

Another interesting possible extension of our results would be to perform a similar analysis for the non-metricity $Q^{\rho}{}_{\mu\nu}$. For $Q^{\rho}{}_{\mu\nu}$ the procedure would be a bit more computationally involved although very similar. Obviously, in this case we do not expect to get the same principal symbol, so we are not guaranteed to have elliptical operators. Therefore, it is not \emph{a priori} automatic to have a unique decomposition, or that it exists at all.
As we have already mentioned, we chose not to consider non-metricity, discarding it from a physical point of view for the following reasons. First, disformation does not easily couple to fermionic spin-$\frac{1}{2}$ matter, though in principle one could be able to couple its hook symmetric part to Rarita-Schwinger spin-$\frac{3}{2}$ fermions. Second, Lagrangians with non-metricity generally propagate a spin-$3$ mode, which is known to give inconsistencies at the level of soft theorems \cite{Weinberg:1965nx}.

A further problem that should be more seriously tackled in the future is the presence of zero modes of the Curtright forms. Indeed, our decomposition is clearly unique only up to zero modes, much like the York decomposition is unique up to (conformal) Killing vectors. However, in this paper, we only scratched the surface of this problem by showing that the zero modes of $\overline{\Delta}_{2,3}$ are always orthogonal to their sources. This is the crucial property that ensures the invertibility of such operators on closed manifolds. Moreover, we displayed the set of equations that need to be solved to explicitly find the kernels of these operators in the case of maximally symmetric spacetimes. It would be valuable to extend such analysis to more general geometries and to try and reinterpret those zero modes as generators of symmetries to be imposed at the level of the action.

The point we would like to stress concerns $1$-loop divergences of the effective action coming from integrating out the torsion tensor on maximally symmetric spaces. Our approach greatly simplifies the computation of such divergences because the contribution of the functional measure reduces to the product of determinants of
minimal operators, schematically of the form $\hat O =-\hat1 \square^k+ \hat E$, for which the Seeley-DeWitt coefficients are known \cite{Barvinsky:1985an}. As an application, we have been able to obtain the logarithmic divergences of the Jacobian associated with our decomposition by using the heat kernel technique for differentially constrained fields.
However, for generic spacetimes we could turn to very general covariant techniques to compute such divergences. For example, we could use the very powerful generalized Schwinger-DeWitt technique developed by Barvinsky and Vilkovisky \cite{Barvinsky:1985an}, which is valid for a very wide class of non-minimal operators. It would be appealing to use both procedures to check if the results match in the appropriate limit.

Let us conclude with some physical insights about the general features of integrating out the torsion around non-symmetric backgrounds at $1$-loop. We expect such radiative effects to naturally generate $R^2$ divergences because these would be absent only if the starting Lagrangian is Weyl invariant with no associated physical vector potential. We also know that the only term which is not Weyl invariant in the Standard Model Lagrangian is the Higgs mass term. Therefore, only the torsion and the Higgs are responsible for generating $R^2$ divergences. The first field is expected to acquire a mass of the order of the Planck scale $10^{19} \, {\rm GeV}$ \cite{Shapiro:2001rz}, while the mass scale of the second is $125 \, {\rm GeV}$. Given the huge mass ratio, we expect $1$-loop radiative corrections due to torsion to be way more relevant during the Planck era. From a phenomenological viewpoint, it would be interesting to study what is the region in the parameter space of the torsion for which the RG-flow of higher derivative gravity is driven towards the Starobinsky Lagrangian \cite{Starobinsky:1980te} in the IR.
%

\section*{Acknowledgment}

The authors would like to thank O. Zanusso for very useful discussions and helpful comments. 

\appendix
\section{Heat kernel formalism for the effective action}

Let us consider an operator $\Delta$ acting on a generic transverse (and possibly traceless) tensor field. Suppose such an operator is minimal and its endomorphism is a multiple of the identity times the Ricci scalar. Then we can surely write $\Delta = \Delta_B + c R$, where $\Delta_B$ is another minimal operator. On maximally symmetric spaces we have $[\Delta_B,cR]=0$, therefore we can split the exponential as ${\rm e}^{-s(-\Delta)}={\rm e}^{s \Delta_B}{\rm e}^{s c R}$. Thus, we can single out the logarithmically divergent part of the effective action as  
\begin{eqnarray}
 &\hspace*{-50px}\Gamma[g]  = - \frac{1}{2} \int_0^\infty \frac{ds}{s} {\rm Tr} K_\Delta(s) = - \frac{1}{2} \int_0^\infty \frac{ds}{s} {\rm Tr} ({\rm e}^{-s(-\Delta)}) = - \frac{1}{2} \int_0^\infty \frac{ds}{s} {\rm e}^{scR} {\rm Tr \, e}^{s\Delta_B} \\\nonumber
 &\hspace*{-50px}\qquad = - \frac{1}{2} \int_0^\infty \frac{ds}{s^3} \left[ 1+scR+\frac{1}{2}s^2 c^2 R^2 + O(s^3) \right] \left[ B_0 + B_2 s + B_4 s^2 + O(s^3) \right] \\\nonumber
 &\hspace*{-50px}\qquad = - \frac{1}{2} \int_0^\infty \left\{\frac{ds}{s} \left[ \frac{1}{2} c^2 R^2 B_0 + c R B_2 + B_4 \right] + \dots  \right\} \, .
\end{eqnarray}
Therefore, if the heat kernel coefficients for the operator $\Delta_B$ are known, it is fairly easy to obtain the coefficients for $\Delta$. Let us assume the operator $\Delta_B$ be of the form
\begin{eqnarray}
\Delta_B =  -\hat{1}g^{\mu\nu}D_{\mu}D_{\nu} + \hat{\rm E}\,,
\end{eqnarray} 
where $\rm \hat{E}$ is an endomorphism on the space of fields, $\hat{1}$ is the identity on the same space and the covariant derivative $D_\mu$ includes possible gauge connections $D_\mu=\nabla_\mu+A_\mu$. Then, the first Seeley-DeWitt coefficients are given by \cite{Percacci:2017fkn}
\begin{eqnarray}
b_0(\Delta_B) =& {\rm tr}\hat{1}\,,\\
b_2(\Delta_B) =& \frac{1}{6}R\, {\rm tr}\hat{1}-{\rm tr}\,\hat{\rm E}\,,\\
b_4(\Delta_B) =& \frac{1}{180}\left(R_{\mu\nu\rho\sigma}R^{\mu\nu\rho\sigma}-R_{\mu\nu}R^{\mu\nu}+\frac{5}{2}R^2+6\square R\right){\rm tr}\hat{1}\label{eq:b4General}\\
	& +\frac{1}{2}{\rm tr}\,\hat{\rm E}^2-\frac{1}{6}R\,{\rm tr}\,\hat{\rm E}+\frac{1}{12}{\rm tr}\,\hat{\Omega}^{\mu\nu}\hat{\Omega}_{\mu\nu}-\frac{1}{6}\square\, {\rm tr}\,\hat{\rm E}\,,\nonumber
\end{eqnarray}
where the box operator only includes the connection on the tangent bundle $\square=g^{\mu\nu}\nabla_\mu \nabla_\nu$, and where $\hat{\Omega}_{\mu\nu}$ is given in terms of the test field $\psi$ as
\begin{eqnarray}
\left[D_{\mu}\,, D_{\nu}\right]\psi = \hat{\Omega}_{\mu\nu} \psi\,.
\end{eqnarray}
The relation between the coefficients $b_{2n}$ and $B_{2n}$ is given in the main text in Eq.\ \eref{eq:bToB}.

While most of the heat kernel coefficients we use in this work are known in the literature, we could not find any explicit computation for the heat kernel of an operator acting on antisymmetric tensors. Therefore, we report here our result for the $b_4$ coefficient found on $S^4$ for the operator $\Delta_A$ in Eq.\ \eref{eq:DeltaA}
\begin{eqnarray*}
b_4\left(-\square+\frac{R}{6}\right)\Big|_{A_{\mu\nu}} = \frac{109}{360}R^2\,,
\end{eqnarray*}
which we found by directly applying Eq.\ \eref{eq:b4General} to the case of maximally symmetric spaces. This result is used in the main text to obtain the result  Eq.\ \eref{eq:b4ATrans} upon subtracting the contributions of $\xi_{\perp}{}^\mu$.

\vspace{1cm}



\end{document}